\documentclass[useAMS,usenatbib]{mnras}
\usepackage[T1]{fontenc}
\usepackage{ae,aecompl}

\usepackage{graphicx}	% Including figure files
\usepackage{amsmath}	% Advanced maths commands
\usepackage{amssymb}	% Extra maths symbols
\usepackage{color}
\usepackage{siunitx}
\newcommand{\Mwd}{\mbox{$M_\mathrm{WD}$}}

\newcommand{\Ha}{\mbox{${\mathrm H\alpha}$}}
\newcommand{\Msun}{\mbox{$\mathrm{M}_{\odot}$}}

\newcommand{\Teff}{\mbox{$T_{\mathrm{eff}}$}}
\newcommand{\logg}{\mbox{$\log g$}}

\newcommand{\Lines}[3]{\Ion{#1}{#2}\,$\lambda\lambda$\,#3}
\newcommand{\Ion}[2]{#1{\,\scriptsize #2}}

\newcommand{\kms}{\mbox{$\mathrm{km\,s^{-1}}$}}
\title[WDMS from LAMOST DR5]{White dwarf-main sequence binaries
from LAMOST: the DR5 catalogue}

\author[J.-J. Ren et al.]
{J.-J.~Ren,$^{1}$\thanks{E-mail: jjren@nao.cas.cn (JJR).}
A.~Rebassa-Mansergas,$^{2,3}$ 
S.~G.~Parsons,$^{4}$ 
X.-W.~Liu,$^{5}$
A.-L.~Luo,$^{1}$
\newauthor
X.~Kong,$^{1,6}$ 
H.-T.~Zhang$^{1}$
\\
$^{1}$Key Laboratory of Optical Astronomy, National Astronomical
Observatories, Chinese Academy of Sciences, Beijing 100012,
P. R. China\\
$^{2}$Departament de F\'isica, Universitat Polit\`ecnica de Catalunya,
c/Esteve Terrades 5, E-08860 Castelldefels, Spain\\
$^{3}$Institute for Space Studies of Catalonia, c/Gran Capit\`a 2--4, Edif. Nexus 201, 08034 Barcelona, Spain\\
$^{4}$Department of Physics and Astronomy, University of Sheffield,
Sheffield S3 7RH, UK\\
$^{5}$South-Western Institute for Astronomy Research, Yunnan
University, Kunming, Yunnan 650091, P. R. China\\
$^{6}$University of Chinese Academy of Sciences, Beijing 10049,
P. R. China\\ }

\date{Accepted XXX. Received YYY; in original form ZZZ}

\pubyear{2018}

\begin{document}
\label{firstpage}
\pagerange{\pageref{firstpage}--\pageref{lastpage}}
\maketitle

% Abstract of the paper
\begin{abstract}
We  present the  data release  (DR)  5 catalogue  of white  dwarf-main
sequence  (WDMS)  binaries  from  the Large  Area  Multi-Object  fiber
Spectroscopic  Telescope (LAMOST).   The catalogue  contains 876  WDMS
binaries, of which 757 are additions to our previous LAMOST DR1 sample
and  357 are  systems that  have not  been published  before. We  also
describe a LAMOST-dedicated  survey that aims at  obtaining spectra of
photometrically-selected  WDMS binaries  from  the  Sloan Digital  Sky
Survey (SDSS)  that are expected  to contain cool white  dwarfs and/or
early type M dwarf companions.  This is a population under-represented
in previous  SDSS WDMS  binary catalogues.   We determine  the stellar
parameters (white dwarf effective  temperatures, surface gravities and
masses, and  M dwarf spectral types)  of the LAMOST DR5  WDMS binaries
and make use of the  parameter distributions to analyse the properties
of the sample.  We find  that, despite our efforts, systems containing
cool white dwarfs remain under-represented.   Moreover, we make use of
LAMOST  DR5  and  SDSS  DR14   (when  available)  spectra  to  measure
the      \Lines{Na}{I}{8183.27,\,8194.81}      absorption      doublet
and/or \Ha\,emission  radial velocities  of our systems.   This allows
identifying  128  binaries   displaying  significant  radial  velocity
variations,  76  of  which  are  new.   Finally,  we  cross-match  our
catalogue with the Catalina Surveys and identify 57 systems displaying
light curve  variations.  These include  16 eclipsing systems,  two of
which are  new, and nine  binaries that are new  eclipsing candidates.
We  calculate  periodograms  from  the photometric  data  and  measure
(estimate) the orbital periods of 30 (15) WDMS binaries.
\end{abstract}


%%%%%%%%%%%%%%%%%%%%%%%%%%%%%%%%%%%%%%%%%%%%%%%%%%

%%%%%%%%%%%%%%%%% BODY OF PAPER %%%%%%%%%%%%%%%%%%

\section{Introduction}

Detached binary stars containing a white dwarf (WD) primary and a main
sequence companion  star are  referred to  as WD-main  sequence (WDMS)
binaries. Depending  on the orbital  separations of the  main sequence
binaries  from which  they  descend, the  previous  evolution of  WDMS
binaries follows two  main different paths. Thus,  in approximately 75
per cent of the cases the  orbital separations are wide enough for the
WD        precursors         to        evolve         as        single
stars                                \citep{deKool1992A&A...261..188D,
Willems2004A&A...419.1057W}. In the remaining  cases the main sequence
binary  components are  close  enough for  the  systems to  experience
dynamically unstable mass transfer  interactions when the more massive
star evolves  into a  red giant or  asymptotic giant.   This generally
results        in         a        common         envelope        (CE)
phase  \citep{Iben1993PASP..105.1373I, Webbink2008ASSL..352..233W}  in
which  friction of  the binary  components  with the  material of  the
envelope leads  to a dramatic  decrease of the binary  separation. The
orbital  energy released  during this  process is  used to  eventually
expel       the        envelope       \citep{Passy2012ApJ...744...52P,
Ricker2012ApJ...746...74R}. Close WDMS binaries  that evolve through a
CE  phase are  known as  post-CE binaries  or PCEBs.  Whilst,
stricly speaking,  a PCEB refers  to any  type of binary  that evolved
through CE evolution -- e.g. hot  subdwarf B stars with close low-mass
main      sequence      companions      \citep{Han2002MNRAS.336..449H,
Han2003MNRAS.341..669H, Heber2016PASP..128h2001H}  -- in this  work we
will only consider a PCEB as a close WDMS binary.

The orbital period distribution of WDMS binaries is
bi-modal. Numerical  simulations predict  the orbital periods  of wide
systems  that did  not evolve  through  a CE  to range  between a  few
hundred  to several  thousand days  \citep{Willems2004A&A...419.1057W,
Camacho2014A&A...566A..86C,  Cojocaru2017MNRAS.470.1442C}.    This  is
observationally  confirmed by  \citet{Farihi2010ApJS..190..275F}.  The
observed   PCEB  orbital   period   distribution   displays  a   clear
concentration               of                systems               at
$\sim$\,8\,h                      \citep{Miszalski2009A&A...496..813M,
Nebot2011A&A...536A..43N}, although  PCEBs with periods as  long as 10
days                  have                  been                  also
identified \citep[e.g.][]{Rebassa2012MNRAS.423..320R}.

After  the CE  phase, PCEBs  evolve  to even  shorter orbital  periods
through  angular  momentum  loss  driven by  magnetic  braking  and/or
gravitational  wave emission.   Therefore, they  may undergo  a second
phase    of    CE     evolution,    leading    to    double-degenerate
WDs    \citep{Rebassa2017MNRAS.466.1575R,   Breedt2017MNRAS.468.2910B,
Kilic2017MNRAS.471.4218K}, or  enter a  semidetached state  and become
cataclysmic        variables       \citep{Gansicke2009MNRAS.397.2170G,
Pala2017MNRAS.466.2855P}          or         super-soft          X-ray
sources    \citep{2015MNRAS.452.1754P}.    Double    degenerate   WDs,
cataclysmic variables  and super-soft X-ray sources  are considered to
be        possible        progenitors         of        type        Ia
supernova   \citep{Langer2000A&A...362.1046L,  Han2004MNRAS.350.1301H,
Wang2010MNRAS.401.2729W,   Wang2012NewAR..56..122W},   which  are   of
important interest for cosmological studies.

Thanks         to         the        Sloan         Digital         Sky
Survey                        \citep[SDSS;][]{York2000AJ....120.1579Y,
Stoughton2002AJ....123..485S}   the    number   of   spectroscopically
confirmed  WDMS binaries  has dramatically  increased during  the last
decade                            \citep{Silvestri2006AJ....131.1674S,
Rebassa2007MNRAS.382.1377R,                 Heller2009A&A...496..191H,
Rebassa2010MNRAS.402..620R,                Rebassa2012MNRAS.419..806R,
Liu2012MNRAS.424.1841L,                    Rebassa2013MNRAS.433.3398R,
Li2014MNRAS.445.1331L, Rebassa2016MNRAS.458.3808R}.  Thus, the current
most     updated    SDSS     WDMS     catalogue    includes     3\,294
binaries\footnote{https://sdss-wdms.org/}, which is by far the largest
and   most   homogeneous   sample   of   compact   binaries   in   the
literature \citep{Rebassa2016MNRAS.458.3808R}.  Among  the 3\,294 SDSS
WDMS,   observational  follow-up   studies   have   resulted  in   the
identification of $\sim$1000  wide binaries that did  not interact and
more     than     200     PCEBs     \citep{Rebassa2007MNRAS.382.1377R,
Schreiber2008A&A...484..441S,            Schreiber2010A&A...513L...7S,
Rebassa2011MNRAS.413.1121R,Rebassa2016MNRAS.458.3808R},  of  which  90
have  available   orbital  periods  \citep{Rebassa2008MNRAS.390.1635R,
Nebot2011A&A...536A..43N,   Rebassa2012MNRAS.423..320R}  and   71  are
eclipsing  \citep{Nebot2009A&A...495..561N, Pyrzas2009MNRAS.394..978P,
Pyrzas2012MNRAS.419..817P,                 Parsons2013MNRAS.429..256P,
Parsons2015MNRAS.449.2194P}.

The superb  sample of SDSS WDMS  binaries has led to  many and diverse
advances in  the field of  astrophysics. Among these we  emphasise the
following:    providing     constraints    on    theories     of    CE
evolution        \citep{Davis2010MNRAS.403..179D,       Zorotovic2010,
DeMarco2011MNRAS.411.2277D}, analysing the Galactic properties of WDMS
binaries through population synthesis studies aimed at reproducing the
ensemble          properties          of         the          observed
population                           \citep{Toonen2013A&A...557A..87T,
Camacho2014A&A...566A..86C,              Zorotovic2014A&A...568A..68Z,
Cojocaru2017MNRAS.470.1442C}, testing the  age-metallicity relation in
the     solar    neighbourhood     \citep{Rebassa2016MNRAS.463.1137R},
identifying    pulsating    low-mass     WDs    in    detached    WDMS
binaries  \citep{Pyrzas2015MNRAS.447..691P},  providing  evidence  for
disrupted   magnetic    braking   \citep{Schreiber2010A&A...513L...7S,
Zorotovic2016MNRAS.457.3867Z},   studying  the   origin  of   low-mass
WDs     \citep{Rebassa2011MNRAS.413.1121R},      investigating     the
age-rotation-activity        relation       of        low-mass       M
stars  \citep{Rebassa2013MNRAS.429.3570R, Skinner2017AJ....154..118S},
constraining  the  critical binary  star  separation  for a  planetary
system  origin   of  WD   pollution  \citep{veras2018MNRAS.473.2871V},
testing the mass-radius  relation of low-mass main  sequence stars and
WDs   \citep{Parsons2012MNRAS.419..304P,   Parsons2012MNRAS.420.3281P,
Parsons2017MNRAS.470.4473P}   and   investigating    the   origin   of
circumbinary             giant              planets             around
PCEBs  \citep{Zorotovic2013A&A...549A..95Z,  Marsh2014MNRAS.437..475M,
Parsons2014MNRAS.438L..91P,             Schleicher2015AN....336..458S,
Bours2016MNRAS.460.3873B}.

However, despite the importance of the  SDSS WDMS binary sample, it is
important  to  emphasise  that   it  suffers  from  serious  selection
effects \citep{Rebassa2016MNRAS.458.3808R}.  This  is because the SDSS
focused mainly on spectroscopic  observations of quasars and galaxies,
which      overlap       in      colour      space       with      hot
WDs  \citep{Richards2002AJ....123.2945R,  Smolcic2004ApJ...615L.141S}.
Hence,  WDMS  containing  cool   WDs  and  early-type  companions  are
under-represented.      In      order       to      overcome      this
issue, \citet{Rebassa2012MNRAS.419..806R} presented a dedicated survey
within SEGUE \citep[The Sloan Extension for Galactic Understanding and
Exploration;][]{Yanny2009AJ....137.4377Y}  that  obtained spectra  for
291 WDMS  binaries containing cool  WDs and/or early  type companions.
Moreover, \citet{Rebassa2013MNRAS.433.3398R}  photometrically selected
3419 WDMS candidates for containing cool WDs.

The Large sky Area Multi-Object fiber Spectroscopic Telescope (LAMOST)
survey is a large-scale  spectroscopic survey which follows completely
different  target selection  algorithms than  SDSS, hence  opening the
possibility for identifying WDMS binaries  that help in overcoming the
selection effects of  the SDSS sample.  \citet{Ren2013AJ....146...82R,
Ren2014A&A...570A.107R} identified a total of 121 LAMOST WDMS binaries
from        the       data        release       (DR)        1       of
LAMOST \citep{Luo2012RAA....12.1243L}.  They found that  the intrinsic
properties of the LAMOST WDMS  binary sample were different from those
of the SDSS population. In  particular, LAMOST WDMS binaries are found
at  considerably  shorter  distances  and  are  dominated  by  systems
containing         early-type        companions         and        hot
WDs  \citep{Ren2014A&A...570A.107R}.  LAMOST  WDMS thus  represent  an
important  addition to  the current  known spectroscopic  catalogue of
SDSS WDMS  binaries. However, the  number of WDMS containing  cool WDs
remains  still  under-represented,  and  the  number  of  LAMOST  WDMS
binaries identified so far is rather small.

Here we present  an updated LAMOST WDMS binary catalogue  based on the
most recent  DR of LAMOST --  DR5, to thus increase  the current small
number  of LAMOST  WDMS  binaries. Moreover,  we  present a  dedicated
LAMOST  survey   for  obtaining  spectra   of  as  many   as  possible
photometrically        selected        SDSS       WDMS        binaries
from \citet{Rebassa2013MNRAS.433.3398R}  that presumably  contain cool
WDs. We measure the stellar parameters  from the LAMOST spectra of all
our identified  objects and  perform a  statistical analysis  of their
intrinsic properties. We also determine the radial velocities (RVs) of
each binary with  the aim of detecting PCEBs.  Finally, we cross-match
our sample  with photometric data  from the Catalina Sky  Survey (CSS)
and           Catalina          Real           Time          Transient
Survey  \citep[CRTS;][]{Drake2009ApJ...696..870D} in  order to  search
for WDMS  binaries displaying  light curve variations  (e.g. eclipsing
systems).

\section{The DR5 of LAMOST}

LAMOST  is a  quasi-meridian reflecting  Schmidt telescope  of $\sim$4
meter  effective  aperture  and  a  field of  view  of  5\,degrees  in
diameter                               \citep{Wang1996ApOpt..35.5155W,
Su2004ChJAA...4....1S,Cui2012RAA....12.1197C}.  LAMOST  is located  in
Xinglong  station of  National Astronomical  Observatories of  Chinese
Academy of  Sciences. Being a dedicated  large-scale survey telescope,
LAMOST uses  4\,000 fibres to  obtain spectra of celestial  objects as
well as sky background and calibration sources in one single exposure.
To that end, LAMOST uses 16 fiber-fed spectrographs each accommodating
250 fibers. Each spectrograph is equipped with two CCD cameras of blue
and red channels  that simultaneously provide blue and  red spectra of
the 4\,000  selected targets, respectively.  The  LAMOST spectra cover
the entire optical wavelength range ($\simeq$\,3700\,--\,9000\AA) at a
resolving power $R$\,$\sim$\,1800.

From 2012 September, LAMOST has  been carrying out a five-year Regular
Survey. Before that,  there was a one-year Pilot Survey  preceded by a
two-year commissioning  phase. The  LAMOST Regular Survey  consists of
two    components    \citep{Zhao2012RAA....12..723Z}:    the    LAMOST
Extra-Galactic Survey  of galaxies (LEGAS)  that aims at  studying the
large scale structure  of the universe, and the  LAMOST Experiment for
Galactic Understanding and Exploration  (LEGUE) that aims at obtaining
millions  of stellar  spectra  in  order to  study  the structure  and
evolution of  the Milky Way \citep{Deng2012RAA....12..735D}.  LEGUE is
sub-divided    into    three    sub-surveys:   the    spheroid,    the
anticentre   \citep{Liu2014IAUS..298..310L,   Yuan2015MNRAS.448..855Y,
Xiang2017MNRAS.467.1890X} and the disc  surveys. The five-year Regular
Survey finished in  June 16th 2017 and the spectra  have been released
internally  to the  Chinese scientific  community through  the DR5  of
LAMOST\footnote{http://dr5.lamost.org/}. The raw spectra are processed
with          the         LAMOST          two-dimensional         (2D)
pipeline \citep{Luo2012RAA....12.1243L, Luo2015RAA....15.1095L}, which
includes   dark   and   bias   subtractions,   cosmic   ray   removal,
one-dimensional (1D) spectral  extraction, merging sub-exposures (note
that each plate  is generally observed consecutively for  a minimum of
three times), and finally, splicing the sub-spectra from the blue- and
red  channels  of  the  spectrographs,  respectively.  The  LAMOST  1D
pipeline is then carried out to perform spectral classification and to
measure the redshift (or radial  velocity) of each spectrum. There are
4\,119  plates observed  in  LAMOST  DR5, thus  including  a total  of
8\,952\,297  spectra, of  which  7\,930\,178 are  stars, 152\,608  are
galaxies,  50\,132  are  quasars,   and  819\,379  are  classified  as
unknowns.

\begin{figure}
	\includegraphics[angle=0,width=80mm]{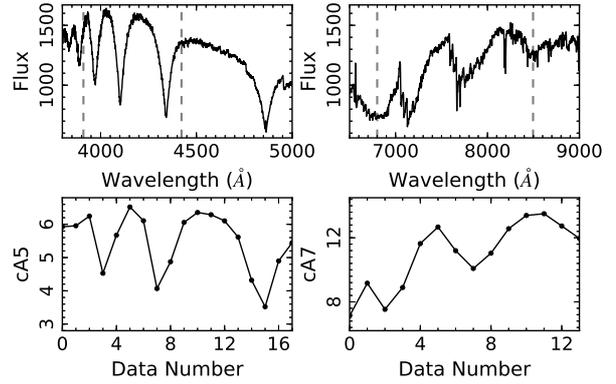} \caption{A
    WT    decomposition   example    of   a    LAMOST   DA/M    binary
    (J161713.52+215517.5). The top panels show the blue (left) and red
    (right) spectrum of the  binary. The vertical grey dashed
    lines  indicate  the  spectral  regions selected  to  perform  the
    wavelet transform (the  3910--4422\AA\, range in the  blue and the
    6800--8496\AA\, range in the red).   The bottom panels display the
    approximation coefficients  (CA) as a  function of data  number of
    points after applying  the WT (5th iteration in  blue, left panel;
    7th iteration in red, right panel).}  \label{fig:wt-example}
\end{figure}

\section{The LAMOST DR5 WDMS binary catalogue}

In  this section  we describe  how  we identify  WDMS binaries  within
LAMOST DR5 and we estimate the completeness of the sample.

\subsection{Identification of WDMS binaries}
\label{s-wt}

In \citet{Ren2014A&A...570A.107R}  we developed  a fast  and efficient
routine            based           on            the           wavelet
transform  \citep[WT;][]{Chui1992wtta.book.....C}   to  identify  WDMS
binaries from  LAMOST DR1. The  analysis unit of  the WT is  the local
flux of the spectrum, i.e.   the selected spectral features.  In other
words, the WT recognises the  spectral features rather than the global
continuum  of  a  given  spectrum.   This is  done  by  decomposing  a
considered  spectral range  (covering the  desired spectral  features)
into  approximation  signals,  often   referred  to  as  approximation
coefficients.   The wavelength  values under  the considered  spectral
region are converted into data  points. This decomposition process can
be  iterated  by  decomposing   the  approximation  coefficients  into
successive  approximation  coefficients  (thus reducing  by  half  the
number of data points in each iteration) until the decomposition level
is  satisfactory. This  occurs when  the spectral  features of  a WDMS
binary spectrum can be identified in the approximation coefficients as
compared  to a  non-WDMS binary  spectrum in  which the  approximation
coefficients are  dominated by  continuum emission and/or  by spectral
features different  from those of WDMS  binaries. The outcome of  a WT
can be  hence considered as  a smoothed  version of the  spectrum (for
comparative purposes).  Because of  its high efficiency in identifying
spectral features, the  WT is very suitable to  identify WDMS binaries
among low signal-to-noise (SN) ratio spectra.

\begin{table*}
\centering
\caption{\label{tab:wdms-sample}
The catalogue of LAMOST DR5 WDMS binaries, including a total of 1\,150
spectra of  876 unique  targets.  Here we  list the  coordinates, SDSS
$ugriz$  magnitudes  (when  available),  LAMOST  spectral  identifiers
(plate,  spectrograph,  fibre  IDs  and  modified  Julian  date  MJD),
signal-to-noise ratio in five different bands (SN$_u$, SN$_g$, SN$_r$,
SN$_i$, SN$_z$) and stellar  parameters (the WD effective temperature,
surface  gravity and  mass,  and  the spectral  type  of  the M  dwarf
companion).  We  also provide Simbad classifications  (when available)
and indicate  if the target  is identified by  the WT (fWT=1,  yes; 0,
no),  if  it is  a  photometrically  selected  SDSS WDMS  binary  that
obtained a LAMOST spectrum (fPhot=1, yes; 0, no), if it is part of the
SDSS DR12 WDMS spectroscopic catalogue (fSDSS=1, yes; 0, no), if it is
a WDMS binary candidate (fCAND=1,  candidates; 0, genuine DA/M) and/or
if it is part of the LAMOST  DR1 WDMS binary catalogue (fDR1=0, not in
DR1; 10, in DR1 genuine DA/M sample, not in DR1 DA/M candidate sample;
11, in  DR1 DA/M candidate sample).   The entire table is  provided in
the electronic version of the paper.  We use `$-$' to indicate that no
magnitude or stellar parameter is available.}
\setlength{\tabcolsep}{2pt}\resizebox{\textwidth}{!}{%
\begin{tabular}{cccccccrrrrrrrrrrrr}
		\hline
          & Jname & fWT	& fPhot & fSDSS & fCAND & fDR1 & RA & Dec & u & Err & g & Err & r & Err & i & Err & z & Err\\
		 & & & & & & & ($^\circ$) & ($^\circ$) & (mag) & (mag) & (mag) & (mag) &  (mag) & (mag) & (mag) & (mag) & (mag) & (mag) \\
		\hline
	&	J000448.23+343627.4	&	1	&	0	&	0	&	0	&	00	&	1.2009999	&	34.6076320	&	$-$	&	$-$	&	$-$	&	$-$	&	$-$	&	$-$	&	$-$	&	$-$	&	$-$	&	$-$ \\
	&	J000612.32+340358.3	&	1	&	1	&	0	&	0	&	00	&	1.5513397	&	34.0661980	&	18.954	&	0.029	&	18.327	&	0.008	&	17.794	&	0.008	&	16.535	&	0.005	&	15.713	&	0.007 \\
	&	J000729.32+023124.5	&	1	&	0	&	0	&	0	&	00	&	1.8721920	&	2.5234990	&	19.912	&	0.042	&	19.092	&	0.010	&	18.379	&	0.008	&	17.494	&	0.007	&	16.937	&	0.011 \\
	&	J001258.26+062617.9	&	1	&	0	&	0	&	0	&	00	&	3.2427542	&	6.4383056	&	18.505	&	0.016	&	17.987	&	0.006	&	18.001	&	0.007	&	17.528	&	0.007	&	17.088	&	0.011 \\
	&	J001733.59+004030.4	&	1	&	0	&	1	&	0	&	00	&	4.3899705	&	0.6751216	&	20.200	&	0.051	&	19.517	&	0.014	&	18.982	&	0.013	&	17.925	&	0.009	&	17.166	&	0.014 \\
	&	J001752.63+332424.9	&	1	&	1	&	0	&	0	&	00	&	4.4693105	&	33.4069250	&	18.582	&	0.015	&	17.625	&	0.005	&	16.519	&	0.004	&	15.616	&	0.004	&	15.069	&	0.005 \\
	&	J001752.63+332424.9	&	1	&	1	&	0	&	0	&	00	&	4.4693105	&	33.4069250	&	18.582	&	0.015	&	17.625	&	0.005	&	16.519	&	0.004	&	15.616	&	0.004	&	15.069	&	0.005 \\
	&	J002237.90+334322.1	&	1	&	0	&	0	&	0	&	10	&	5.6579420	&	33.7228090	&	19.548	&	0.033	&	17.361	&	0.005	&	15.985	&	0.003	&	14.590	&	0.004	&	13.858	&	0.004 \\
	&	J002407.31+054856.4	&	1	&	1	&	0	&	0	&	00	&	6.0304700	&	5.8156830	&	17.068	&	0.008	&	16.393	&	0.004	&	15.851	&	0.004	&	14.808	&	0.004	&	14.128	&	0.004 \\
	&	J002633.13+390904.0	&	1	&	1	&	0	&	0	&	00	&	6.6380824	&	39.1511210	&	16.348	&	0.006	&	15.905	&	0.003	&	16.011	&	0.004	&	15.831	&	0.004	&	15.327	&	0.005 \\
	&	J002633.14+390904.0	&	1	&	1	&	0	&	0	&	00	&	6.6385030	&	39.1508170	&	16.348	&	0.006	&	15.905	&	0.003	&	16.011	&	0.004	&	15.831	&	0.004	&	15.327	&	0.005 \\
\end{tabular}}
\setlength{\tabcolsep}{6pt}
\resizebox{\textwidth}{!}{%
\begin{tabular}{cccccrrrrrrrrrrrcl}
		\hline
         	&	MJD	&	Plate	&	spID	&	fiberID	&	SN$_u$	&	SN$_g$	&	SN$_r$	&	SN$_i$	&	SN$_z$	&	\Teff$_\mathrm{WD}$	&	Err	&	\logg$_\mathrm{WD}$	&	Err	&	\Mwd	&	Err	&	Sp	&	Simbad \\
		   & & & & & & & & & & (K) & (K) & (dex) & (dex) & (M$_{\sun}$) & (M$_{\sun}$) & & \\
		\hline
	&	56948	&	M31001N35M2         &	05	&	156	&	8.66	&	33.80	&	30.05	&	49.00	&	32.49	&	16717	&	245	&	8.02	&	0.06	&	0.63	&	0.03	&	4	&	$-$	\\
	&	56948	&	M31001N35M2	        &	01	&	139	&	3.92	&	12.66	&	22.09	&	47.00	&	50.04	&	28389	&	1448	&	7.87	&	0.23	&	0.58	&	0.11	&	4	&	$-$	\\
	&	55893	&	F9302	            &	01	&	235	&	1.72	&	2.52  	&	4.11 	&	7.63	    &	6.58 	&	$-$	    &	$-$	&	$-$	&	$-$	&	$-$	&	$-$	&	1	&	$-$	\\
	&	56602	&	EG001605N080655M01	&	02	&	066	&	1.52	&	3.01 	&	2.25 	&	3.48 	&	3.23 	&	$-$	    &	$-$	&	$-$	&	$-$	&	$-$	&	$-$	&	4	&	$-$	\\
	&	56978	&	EG001639N015102M01	&	05	&	141	&	1.52	&	5.07 	&	7.31	    &	16.86	&	14.32	&	$-$	    &	$-$	&	$-$	&	$-$	&	$-$	&	$-$	&	3	&	DA+dM	\\
	&	56952	&	M31005N35M1	        &	02	&	120	&	2.44	&	13.88	&	35.84	&	71.58	&	60.63	&	$-$	    &	$-$	&	$-$	&	$-$	&	$-$	&	$-$	&	1	&	$-$	\\
	&	57717	&	M31007N33B2	        &	10	&	109	&	3.66	&	16.36	&	34.31	&	71.88	&	63.27	&	29727	&	1312	&	8.34	&	0.25	&	0.82	&	0.14	&	2	&	$-$	\\
	&	56255	&	VB007N33V1	        &	03	&	085	&	2.23	&	3.49	    &	2.78 	&	1.72	    &	3.15 	&	$-$	    &	$-$	&   $-$	&	$-$	&	$-$	&	$-$	&	$-$	&	DA+M	\\
	&	56621	&	EG002616N034932B01	&	11	&	020	&	8.25	&	27.85	&	40.10	&	86.82	&	82.41	&	14899	&	395	&	8.41	&	0.08	&	0.85	&	0.05	&	3	&	$-$	\\
	&	57280	&	M31007N36M1	        &	11	&	163	&	1.80	&	3.81	    &	4.02 	&	4.78	    &	4.40	    &	$-$	    &	$-$	&	$-$	&	$-$	&	$-$	&	$-$	&	9	&	DA4.6	\\
	&	56911	&	HD002951N381926B01	&	15	&	227	&	4.28	&	9.16 	&	5.54 	&	4.01 	&	3.84	    &	$-$	    &	$-$	&	$-$	&	$-$	&	$-$	&   $-$	&	$-$	&	DA4.6	\\
		\hline
\end{tabular}}
\end{table*}
%\end{landscape}

\begin{figure}
	\includegraphics[angle=0,width=80mm]{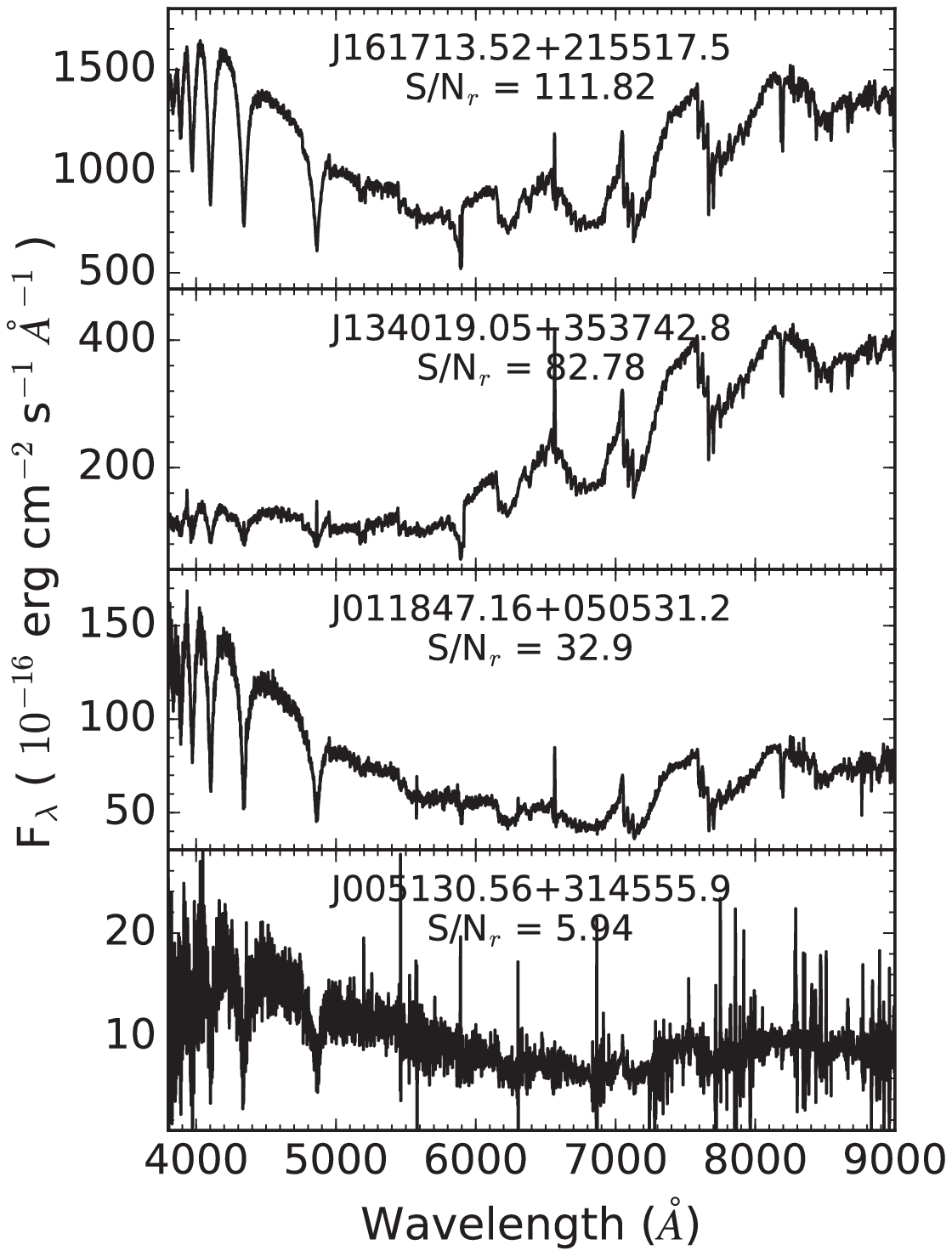} \caption{Example
        spectra  of  LAMOST DA/M  binaries  identified  by the  WT  in
        descendant    order     of    SN     ratio    (in     the    r
        band).}  \label{fig:spec-example}
\end{figure}

WDMS binaries containing  a DA (hydrogen-rich) WD and  a low-mass main
sequence star have two obvious  spectral features: the Balmer lines in
the blue band  arising from the WD and the  molecular absorption bands
in the red  that are typical of  M dwarfs. We hence applied  the WT to
the following spectral regions: the  3910--4422\AA\, range in the blue
band  that covers  the H$\beta$  to H$\epsilon$  Balmer lines  and the
6800--8496\AA\, range which  covers a large number  of TiO and
VO molecular bands.  By choosing these two spectral features we focus
our search on  the identification of WDMS binaries containing  a DA WD
and a M dwarf companion (hereafter DA/M binaries; see an example of WT
applied          to         a          DA/M         binary          in
Figure\,\ref{fig:wt-example}).   The  main   reason  for  this
choice is  that DA/M are by  far the most common  among WDMS binaries,
including            $\sim$\,80\,per\,cent           of            all
systems  \citep{Rebassa2016MNRAS.458.3808R}. This  is  because of  the
following two reasons. First, DA WDs comprise $\sim$\,80\,per\,cent of
all  single WDs  \citep{Kepler2015MNRAS.446.4078K},  hence  it is  not
surprising that they are the most common WDs in binaries too.  Second,
because of  selection effects,  main sequence  stars of  spectral type
earlier  than M  generally  outshine  the WDs  in  the optical,  which
implies only  a handful of WDs  with late K-type companions  have been
identified  \citep{Rebassa2016MNRAS.458.3808R}.   We will  pursue  the
identification of LAMOST  WDMS binaries harboring other  WD types (DB,
DQ, DZ, etc)  as well as late K-type companions  elsewhere. Of course,
in order for our WT to  efficiently identify DA/M binaries we require
the two components (DA WD plus M dwarf) to be visible in the spectrum.
This implies that  systems in which one of the  two components clearly
dominates the  spectral energy  distribution will  be harder  (or even
impossible)  to detect.   We note  however that  this issue  makes the
detection   of   such   systems   to  be   very   difficult   by   any
method \citep[see for example][]{Rebassa2010MNRAS.402..620R}.

After  applying  the WT  method  to  the  $\sim$9 million  LAMOST  DR5
spectra, we obtained an initial  list of 29\,269 WDMS binary candidate
spectra that we visually inspected.  This resulted in 776 spectra (579
unique systems)  that we considered  as genuine DA/M binaries,  and 47
spectra (46  unique systems)  that we  catalogued as  DA/M candidates.
The majority  of spectra selected by  the WT that we  did not consider
DA/M binaries corresponded to single MS stars and to the superposition
in   the    line   of    sight   of    two   main    sequence   stars.
Table\,\ref{tab:wdms-sample} lists all the  genuine and candidate DA/M
binaries  identified  by   the  WT,  where  we   also  include  LAMOST
spectroscopic   identifiers  and   modified   Julian   dates  of   the
observations.  In Figure\,\ref{fig:spec-example}  we show four example
spectra of LAMOST DA/M binaries of different SN ratio.

\subsection{Completeness of the LAMOST DR5 DA/M binary sample}
\label{s-compl}

In \citet{Ren2014A&A...570A.107R}  we showed  the WT method  is highly
efficient ($\sim$90  per cent) in  identifying DA/M binaries  and that
the spectra it  fails to identify are either clearly  dominated by one
of the stellar  components and/or are associated to very  low SN ratio
($<5$). In order  to estimate the completeness of the  LAMOST DR5 DA/M
binary catalogue (i.e. the number  of DA/M binaries observed by LAMOST
that  the WT  successfully  identified), we  cross-matched the  entire
$\sim$9 million spectroscopic  data base of LAMOST with  the SDSS DR12
spectroscopic  catalogue of  WDMS binaries.  We found  447 objects  in
common (583 LAMOST spectra).

111  (153 spectra)  of the  447  objects (583  spectra) were  non-DA/M
binaries  that the  WT failed  to  identify for  obvious reasons:  the
spectral features of these WDMS  binaries were different from those of
DA/M  binaries.   In order  to  compile  a  catalogue as  complete  as
possible, we added  the 111 objects (105 genuine WDMS,  143 spectra; 6
WDMS binary candidates, 7 spectra) to our list.

The remaining 330  systems were DA/M binaries (424  spectra), 233 (312
spectra) of which were identified by  the WT and 97 (112 spectra) were
missed. This  translated into a  completeness of $\sim$70 per  cent, a
value which is  considerably lower than the $\sim$90  per cent claimed
by \citet{Ren2014A&A...570A.107R}. We  repeated the exercise excluding
all spectra of SN ratio below 5  and found the number of DA/M binaries
identified and  missed by  the WT  were 156 (188  spectra) and  37 (42
spectra), respectively, i.e.  a completeness of $\sim$80  per cent. We
visually inspected the 42 missed spectra  and realised that 19 of them
were clearly  dominated by the flux  contribution of the WD  and 15 by
the  flux  contribution  of  the  M dwarf.  We  also  found  5  broken
spectra. Excluding  the broken spectra  and those dominated by  one of
the  stellar components,  the completeness  increased to  $\sim$98 per
cent.  We added  the  97  objects (93  genuine  WDMS,  108 spectra;  4
candidate WDMS,  4 spectra) the  WT missed to  our list of  LAMOST DR5
WDMS binaries.

We conclude  the WT is  highly efficient at identifying  DA/M binaries
and that  the LAMOST DR5  DA/M binary  catalogue is $\sim$98  per cent
complete. However,  the completeness drops  to $\sim$70 per  cent when
considering spectra clearly dominated by  the flux contribution of one
of  the  stellar  components  and/or  spectra of  very  low  SN  ratio
($<$5).  Unfortunately,  identifying  DA/M   binary  spectra  of  such
characteristics         is         challenging         for         any
method \citep{Rebassa2010MNRAS.402..620R}.

\section{A dedicated LAMOST survey of photometrically selected SDSS WDMS binaries}
\label{s-phot}

The spectroscopic catalogue  of SDSS WDMS binaries  is severely biased
against systems containing cool WDs and/or early type companions. This
is mainly because SDSS dedicated most of its efforts to obtain spectra
of  quasars and  galaxies, which  overlap  in colour  space with  WDMS
binaries containing hot ($\ga$10\,000--15\,000 K) WDs and/or late type
($>$M3--4)   companions.   In   order  to   overcome  this   selection
effect,   \citet{Rebassa2013MNRAS.433.3398R}    developed   a   colour
selection criteria that combines  optical (SDSS $ugriz$) plus infrared
(UKIRT Infrared  Sky Survey  $yjhk$, Two Micron  All Sky  Survey $JHK$
and/or Wide-Field  Infrared Survey  Explorer $w_1 w_2$)  magnitudes to
photometrically  select 3\,419  WDMS binary  candidates for  harboring
cool WDs and/or dominant (M dwarf) companions in their spectra. Visual
inspection of  567 photometric candidates with  available SDSS spectra
allowed   \citet{Rebassa2013MNRAS.433.3398R}  to   conclude  that   84
per\,cent of entire the photometric  sample are expected to be genuine
WDMS binaries, and  that 71 per\,cent should contain  WDs of effective
temperatures  below  10\,000--15\,000\,K  and M  dwarf  companions  of
spectral types concentrated at $\sim$M2--3.

In this section we present a dedicated LAMOST survey that so far has
obtained spectra of 622 photometrically selected SDSS WDMS binaries.

\begin{figure}
	\includegraphics[angle=0,width=\columnwidth]{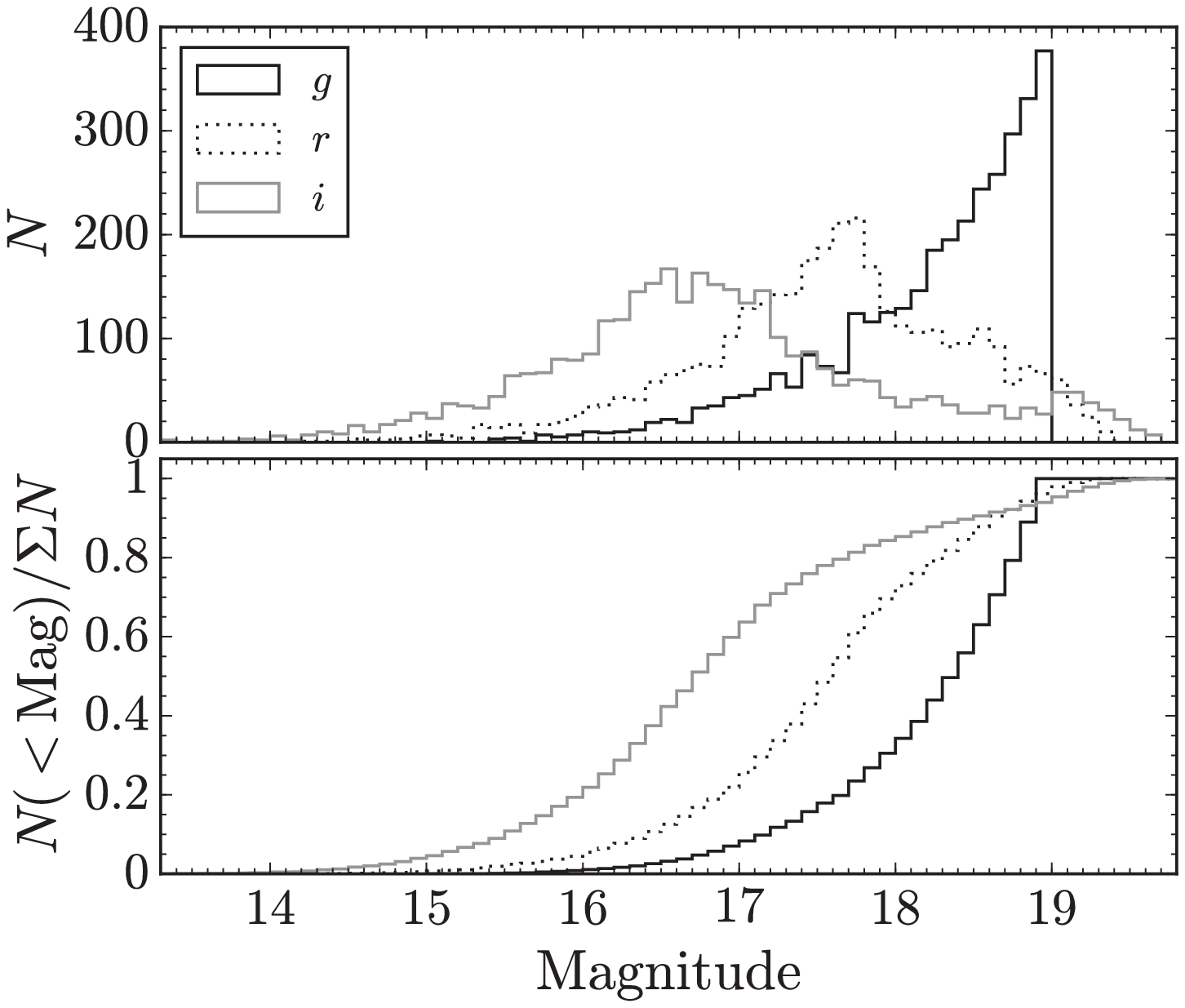} \caption{SDSS
    $gri$  magnitude  distributions   of  the  3\,419  photometrically
    selected                       WDMS                       binaries
    of \citet{Rebassa2013MNRAS.433.3398R}.}  \label{fig:mag-hist}
\end{figure}

\subsection{The LAMOST observations}

\begin{figure*}
	\includegraphics[angle=0,width=0.9\textwidth]{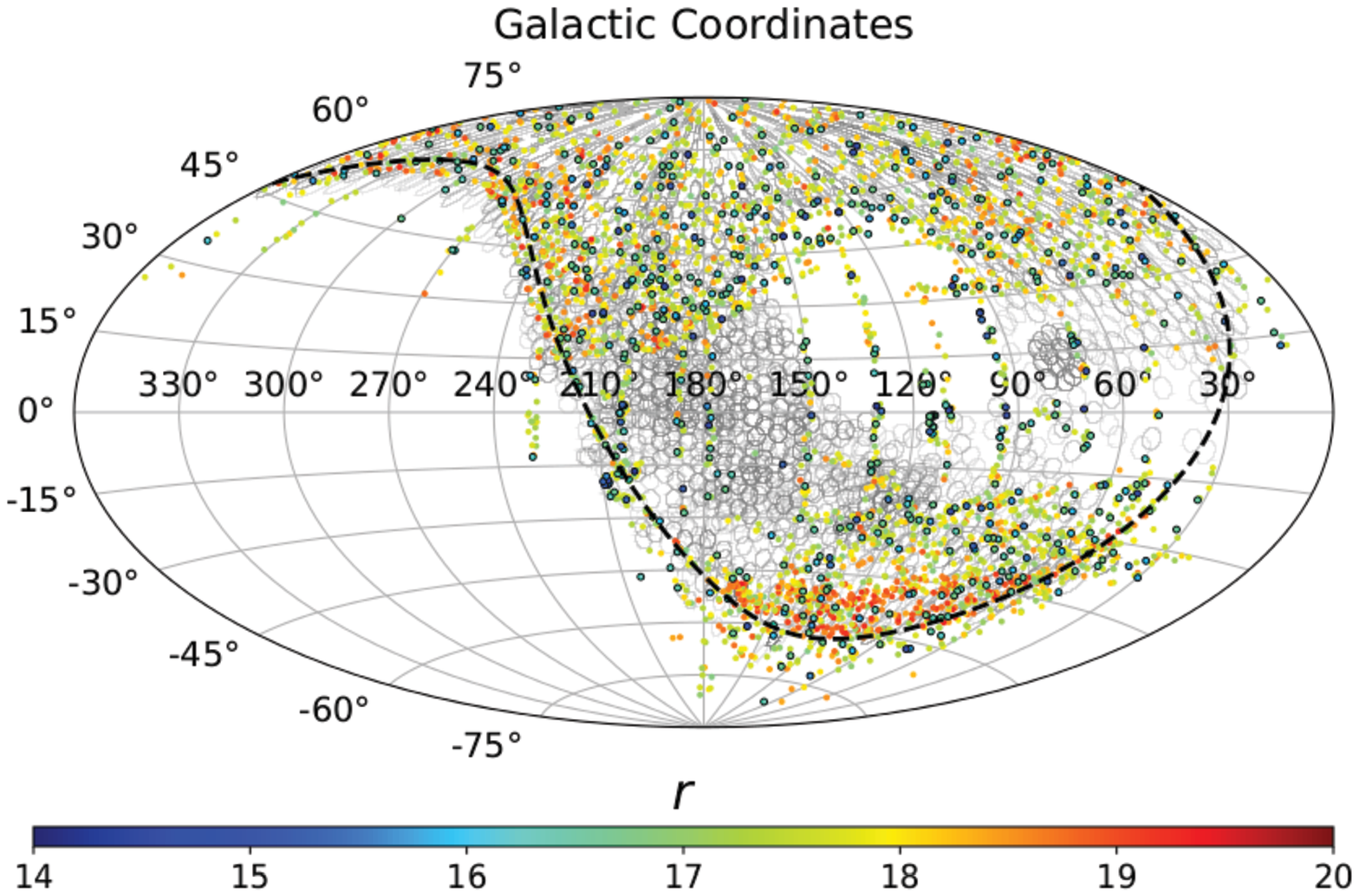} \caption{Galactic
    coordinates of the  LAMOST DR5 plates (gray open  circles) and the
    photometrically       selected       SDSS      WDMS       binaries
    of \citet{Rebassa2013MNRAS.433.3398R}  (solid coloured  dots; each
    colour is associated to a given $r$ magnitude according the colour
    bar displayed in the bottom of  the figure). The coloured dots with
    black   edges  illustrate   those  WDMS   binaries  with   $r\  <$
    16.8\,mag.}  \label{fig:dr5-dist}
\end{figure*}

Since 2014, the LAMOST target selection algorithm has incorporated the
photometrically        selected         WDMS        binary        list
of  \citet{Rebassa2013MNRAS.433.3398R} with  the aim  of obtaining  as
many    WDMS   binary    spectra    as    possible.    As    mentioned
in \citet{Luo2015RAA....15.1095L}, the LAMOST observations are carried
out  using  four  different  modes, i.e.   the  selected  objects  are
observed using  different plates  according to their  magnitudes: very
bright plates (VB, $r \leq$  14\,mag), bright plates (B, 14\,mag $<r<$
16.8\,mag),  medium-bright   plates  (M,   16.8\,mag  $\leq   r  \leq$
17.8\,mag) and faint plates (F, $r  >$ 17.8\,mag). LAMOST DR5 has made
use of  $\sim$\,4\,000 plates, of  which 80 per\,cent are  VB/B plates
(46 per\,cent are VB plates), i.e.   most of the observed targets have
$r$ magnitudes  below 16.8.   Fig\,\ref{fig:mag-hist} shows  the $gri$
magnitude  distribution of  the 3\,419  photometrically selected  WDMS
binaries of \citet{Rebassa2013MNRAS.433.3398R},  where one can clearly
see  the  number of  systems  with  $r<$  16.8  mag is  low  ($\sim$17
per\,cent). Moreover, it is important to emphasize that, on occasions,
the coordinates  of the  LAMOST plates  and the  WDMS binaries  do not
overlap  (see Fig\,\ref{fig:dr5-dist}).   Overall,  this implies  that
only a  small fraction of  photometrically selected WDMS  binaries has
been so far observed by LAMOST.

\subsection{The success rate of the SDSS photometric sample}

The LAMOST observations above described  resulted in a total number of
872      spectra       of      622      unique       sources      from
the \citet{Rebassa2013MNRAS.433.3398R}'s  list. We  visually inspected
the spectra and identified 298 genuine WDMS binaries (436 spectra) and
41  (57 spectra)  that we  considered as  WDMS binary  candidates. The
remaining sources were cataclysmic variables (7; 13 spectra), single M
dwarfs (161;  214 spectra),  quasars (23;  36 spectra),  main sequence
stars (16; 26 spectra), single WDs  (4; 4 spectra), main sequence plus
main sequence superpositions (11; 20 spectra), and unknown objects due
to the bad quality of their spectra (61; 66 spectra).

\begin{figure}
	\includegraphics[angle=-90,width=\columnwidth]{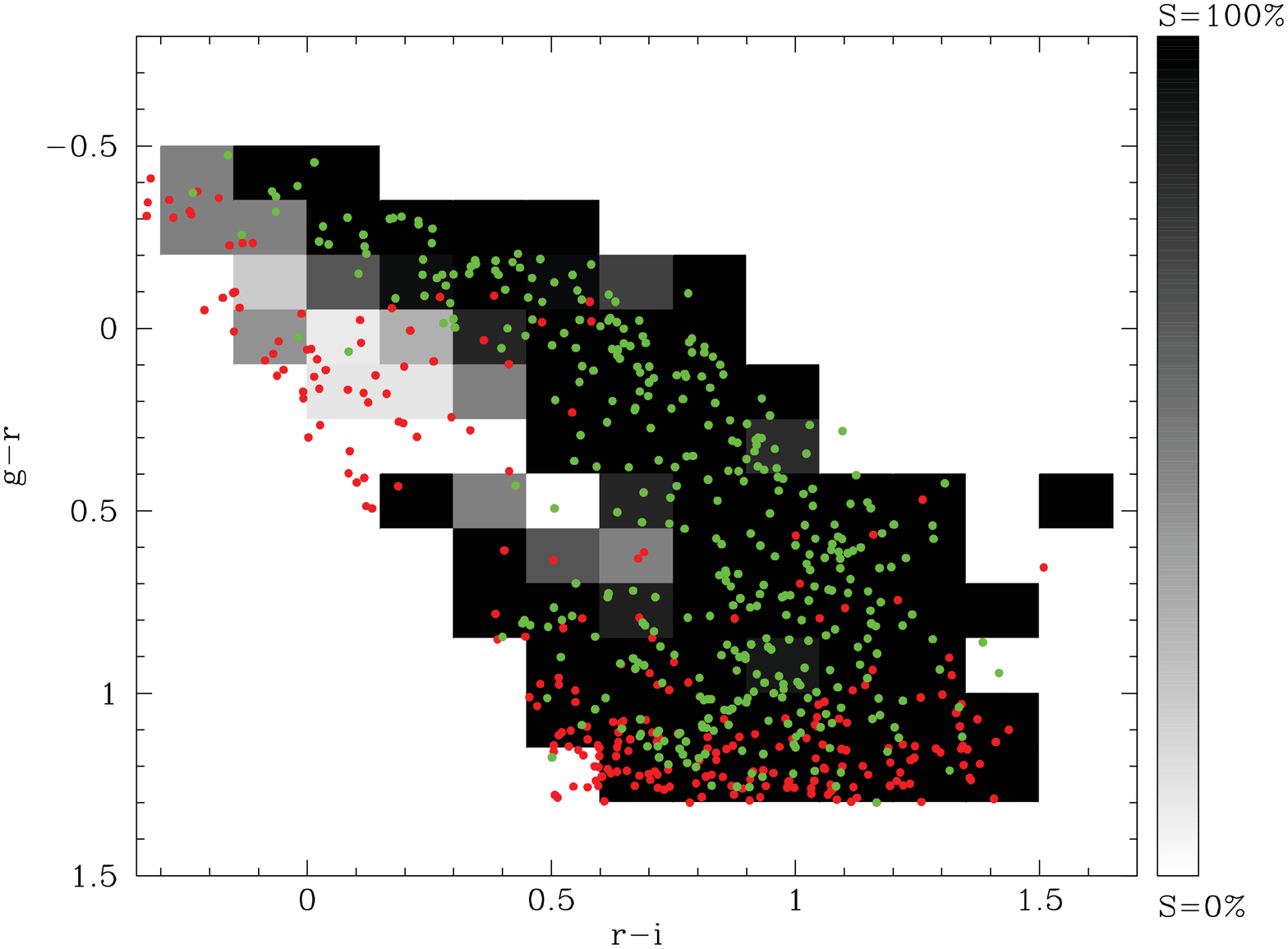} \caption{The
    estimated success  rate for selecting genuine  WDMS binaries based
    on      the      colour     selection      criteria      developed
    by \citet{Rebassa2013MNRAS.433.3398R} (black regions represent 100
    per  cent of  success rate,  white regions  0 per  cent). The  873
    spectra observed by LAMOST DR5  are indicated as solid dots (green
    for       genuine        WDMS       binaries,        red       for
    contaminants).}  \label{fig:phot}
\end{figure}

If we exclude the  61 unknown sources we find that 60  per cent of the
photometric candidates are indeed genuine WDMS binaries or WDMS binary
candidates. This value  is low compared to the 84  per cent of success
rate  claimed  by   \citet{Rebassa2013MNRAS.433.3398R}  for  the  same
sample.      However,     it     is     important     to     emphasise
that \citet{Rebassa2013MNRAS.433.3398R} showed the success rate varies
from 40 per cent up to 100 per cent depending on colour space and that
the 84  per cent  value represents  an average  success rate  over the
entire    colour   space.     This    can   clearly    be   seen    in
Figure\,\ref{fig:phot}, where  we illustrate the success  rate in form
of a density map  in the $g-r$ vs. $r-i$ colour  space. In this figure
we also  show the location  of the  genuine WDMS binaries  observed by
LAMOST DR5 (green solid dots)  and those that we consider contaminants
(i.e. non WDMS binaries). It becomes obvious that the contaminants are
either located in  areas of expected low success  rate or concentrated
at $g-r <$\,1. At such red  colours the WD contribution in the optical
spectrum is  expected to be  rather low,  which makes it  difficult to
judge  simply by  eye  whether  or not  these  objects  are real  WDMS
binaries or  single MS stars.   Our visual inspection decided  for the
latter, although the former cannot be ruled out.

We checked  how many of  the 298  genuine WDMS binaries  identified in
this  section were  also found  by the  WT and  realised that  only 30
objects (a  total of  49 spectra)  were missing. We  made use  of this
result  to derive  (in an  alternative  way) the  completeness of  the
LAMOST  DR5 WDMS  binary  catalogue,  which results  in  this case  in
$\sim$\,90 per  cent.  This value  is slightly  lower than the  98 per
cent  we  obtained  before (see  Section\,\ref{s-compl}),  however  it
confirms the WT  is an efficient tool to identify  WDMS binaries among
large spectroscopic  samples.  We also  found that 37 (52  spectra) of
the 41 (57 spectra) WDMS binary  candidates were not identified by the
WT. Visual  inspection revealed most  of these candidate  spectra were
dominated by  the flux of  the M dwarf,  some were broken,  and others
were of low SN ratio. Hence, it  is not surprising that the WT was not
successful at identifying these spectra.

Among the 30 objects the WT  missed (49 spectra), 14 (25 spectra) were
found cross-matching  the entire LAMOST  data base with the  SDSS DR12
WDMS binary catalogue (see Section\,\ref{s-compl}). Hence, we added to
our list the remaining 16 WDMS binaries (24 spectra). In the same way,
among the 37 WDMS binary candidates (52 spectra) the WT missed, 10 (11
spectra) were  identified in  Section\,\ref{s-compl}.  Thus,  we added
the remaining 27 (41 spectra) WDMS binary candidates to our LAMOST DR5
catalogue.

\section{The final catalogue of LAMOST DR5 WDMS binaries}

The  WT identified  579 genuine  (776  spectra) and  46 candidate  (47
spectra) WDMS binaries (Section\,\ref{s-wt}). To this list we added 97
(112  spectra) DA/M  binaries  (93 genuine  binaries,  108 spectra;  4
candidates, 4  spectra) and 111  (150 spectra) non DA/M  binaries (105
genuine non  DA/M, 143 spectra; 6  non DA/M candidates, 7  spectra) by
cross-matching the  SDSS DR12  WDMS catalogue  with the  entire LAMOST
spectroscopic data base (Section\,\ref{s-compl}). Finally, we added 16
(24 spectra)  genuine and 27  (41 spectra) candidate DA/M  binaries by
visually  inspecting the  LAMOST spectra  of 622  SDSS photometrically
selected targets (Section\,\ref{s-phot}). This brings the total number
of genuine  LAMOST DR5 WDMS  binaries to  793 (1\,051 spectra)  and 83
WDMS  binary  candidates  (99  spectra), i.e.  876  objects  and  1150
spectra.     The      entire     catalogue     is      included     in
Table\,\ref{tab:wdms-sample}.

We  compared  our  LAMOST  DR5  catalogue  to  the  one  we  presented
in  \citet{Ren2014A&A...570A.107R} and  found  that  119 objects  (194
spectra) were already identified in the  DR1 of LAMOST. Hence, the DR5
catalogue contains 757  (956 spectra) new entries.  Moreover, we found
that 400  (509 spectra)  objects were already  discovered in  the SDSS
DR12  catalogue of  \citet{Rebassa2016MNRAS.458.3808R}, which  implies
357 (447 spectra) are new systems that have not been published before.

\section{Characterization of the LAMOST DR5 WDMS binary catalogue}

In this section we measure the WD stellar parameters and determine the
M dwarf  spectral subclass of all  DA/M binaries that are  part of our
new LAMOST DR5 WDMS binary  catalogue. We also describe the properties
of the new DR5 sample by analysing the parameter distributions.

\begin{figure}
\centering
	\includegraphics[angle=-90,width=\columnwidth]{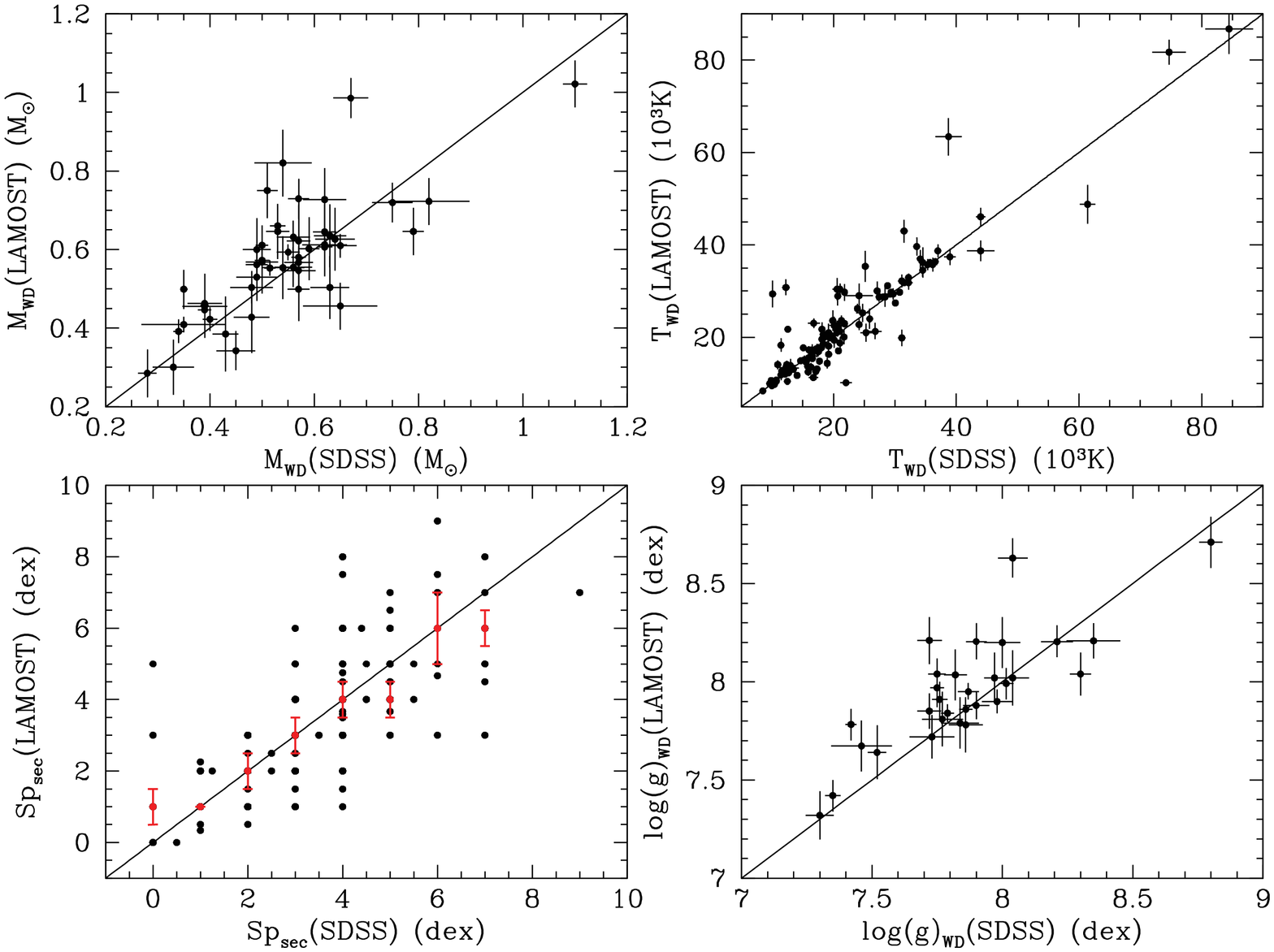} \caption{Comparison
    between the WD parameters  (effective temperature, surface gravity
    and  mass) and  M dwarf  spectral subclass  derived from  SDSS and
    LAMOST spectra of common DA/M binaries.}  \label{fig:param}
\end{figure}

\subsection{Stellar parameter determinations}
\label{s-param}

We    determined     the    stellar    parameters     following    the
decomposition/fitting                 routine                described
by \citet{Rebassa2007MNRAS.382.1377R}.  The first  step was to fit the
DA/M  binary  spectrum with  a  two-component  model  using a  set  of
observed  M  dwarf  and  WD templates,  which  allowed  obtaining  the
spectral type of the M dwarf.  Then, the best-fit M dwarf template was
subtracted   from   the   DA/M   binary  spectrum   and   the   Balmer
lines (H$\beta$  to H$\epsilon$) of the  residual WD spectrum
were    fitted     using    a    grid    of     DA    WD    atmosphere
models \citep{Koester2010MmSAI..81..921K}  to derive the  WD effective
temperature  (\Teff)  and surface  gravity  (\logg).   Given that  the
equivalent widths  of the Balmer lines  reach a maximum near  \Teff\ =
13\,000\,K  (with  the  exact  value   being  a  function  of  \logg),
the \Teff\  and \logg\  determined from Balmer  line profile  fits are
subject  to a  degeneracy. That  is, fits  of similar  quality can  be
achieved  on either  side  of  the temperature  at  which the  maximum
equivalent width  is occurring. We  broke this degeneracy  fitting the
entire WD residual spectrum (continuum plus Balmer lines).

It has  been shown  that one-dimensional  WD model  atmosphere spectra
such as those used in this  work yield overestimated \logg\ values for
WDs          of          effective         temperatures          below
$\sim$12\,000\,K                    \citep{Koester2009JPhCS.172a2006K,
Tremblay2011A&A...531L..19T}.  Therefore, we  applied the  corrections
of   \citet{Tremblay2013A&A...559A.104T},    which   are    based   on
three-dimensional  WD model  atmosphere spectra,  to our  WD parameter
determinations. We then  interpolated the \Teff\ and  \logg\ values in
the cooling  sequences of \citet{Renedo2010ApJ...717..183R}  to derive
the WD masses.  In cases where more than one  spectrum per target were
available    we    averaged    the    stellar    parameters    (namely
the \Teff , \logg\ and  mass of the WD and the spectral  type of the M
dwarf).     The     stellar     parameters     are     provided     in
Table\,\ref{tab:wdms-sample} for each individual spectrum.

\begin{figure}
  \begin{center} \includegraphics[width=\columnwidth]{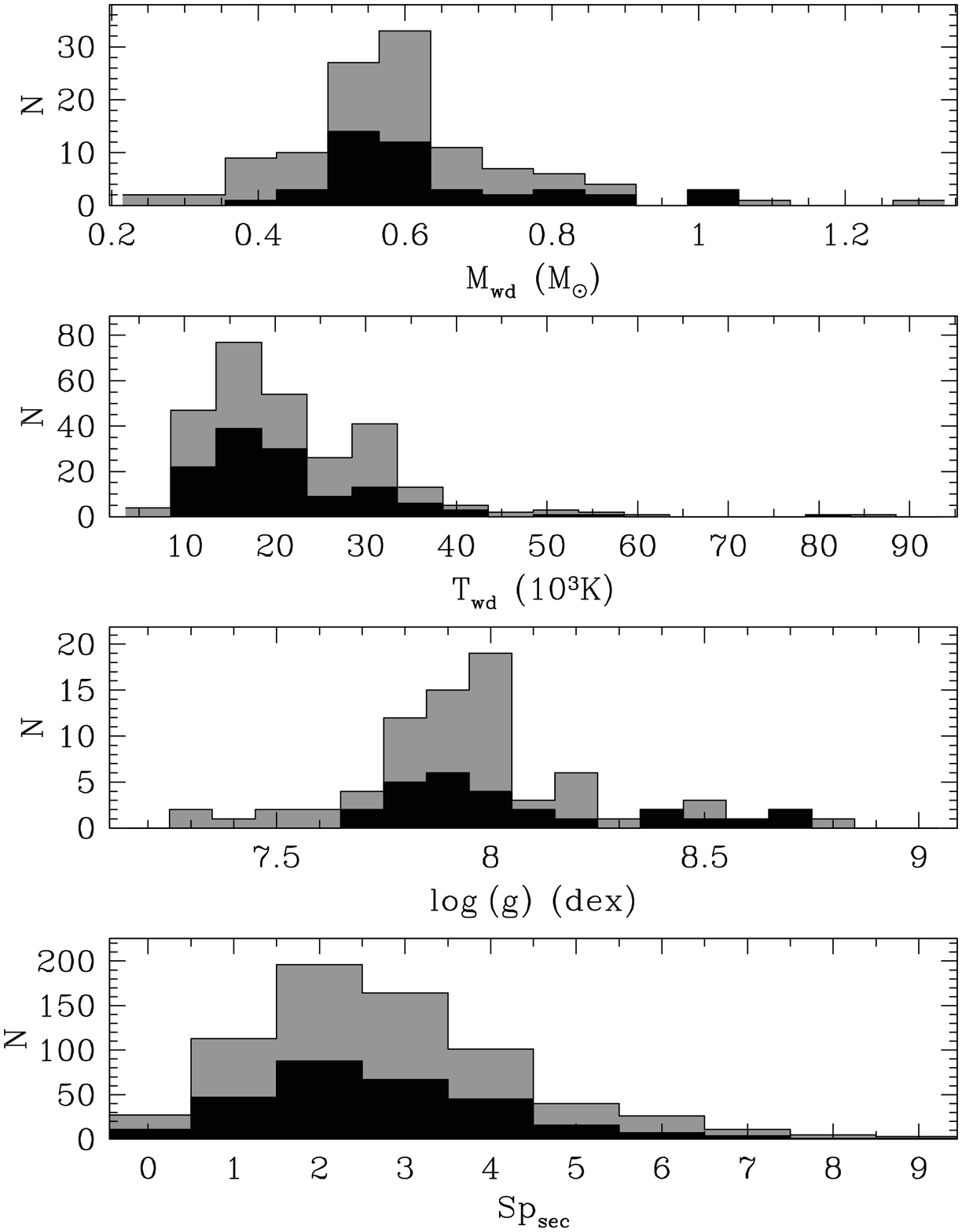} \caption{\label{fig:histo}
    Parameter  distributions   (WD  mass,  top  panel;   WD  effective
    temperature, top-middle  panel; WD surface  gravity, bottom-middle
    panel; secondary M  dwarf spectral subtype, bottom  panel) for the
    entire  LAMOST  DR5  DA/M  binary catalogue  (gray)  and  for  the
    sub-sample of objects that  were photometrically selected from the
    SDSS and observed by LAMOST (black).}  \end{center}
\end{figure}

\begin{figure*}
  \begin{center}                           \includegraphics[angle=-90,
  width=\columnwidth]{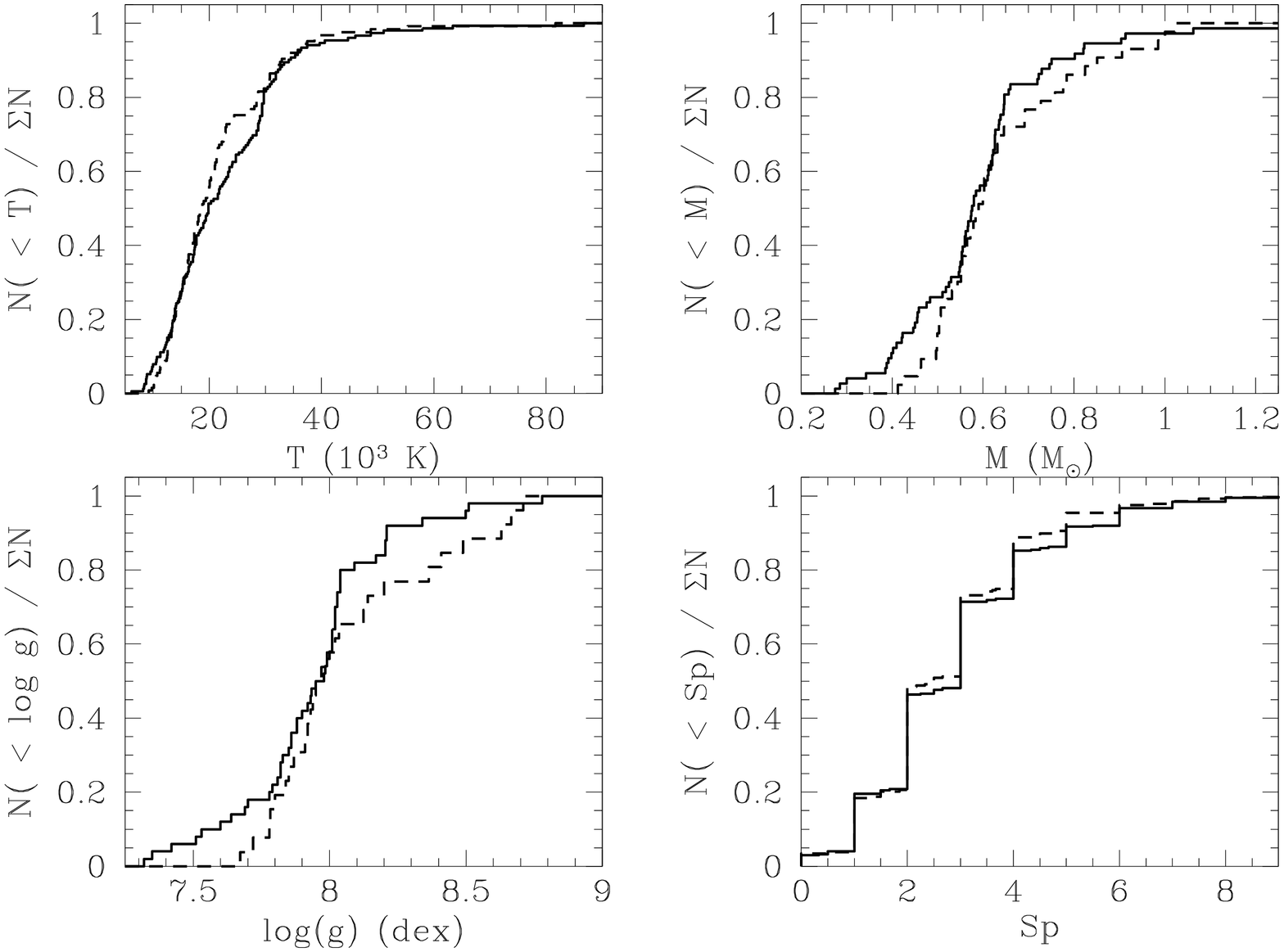}              \includegraphics[angle=-90,
  width=\columnwidth]{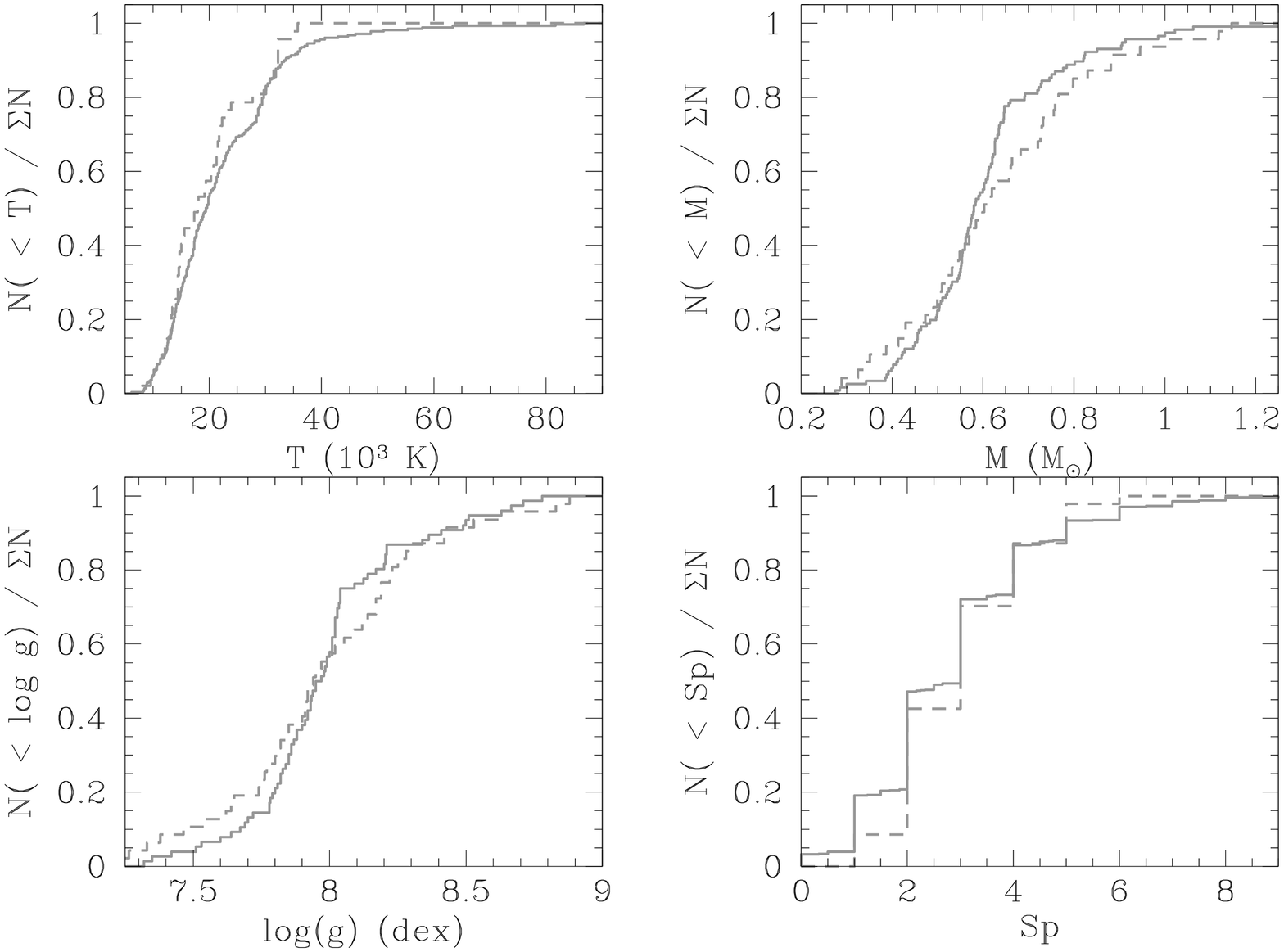}    \caption{\label{fig:ks}    Cumulative
  parameter distributions of the  SDSS photometrically selected sample
  observed  by  LAMOST  (black  dotted  lines;  left  and  middle-left
  panels),  the   complete  DR5  LAMOST  sample   excluding  the  SDSS
  photometrically-selected   ones  (black   solid   lines;  left   and
  middle-left      panels),     the      LAMOST     DR1      catalogue
  of  \citet{Ren2014A&A...570A.107R}  (gray  dotted lines;  right  and
  middle-right panels)  and the  complete LAMOST DR5  catalogue (solid
  gray lines; right and middle-right panels).}  \end{center}
\end{figure*}

Given  that a  large number  of LAMOST  DA/M binaries  have also  been
observed  by  SDSS,  we  decided to  compare  the  stellar  parameters
determined here with those obtained  fitting the SDSS spectra. We note
that we applied  the same fitting routine to both  the LAMOST and SDSS
spectra. The result  can be seen in  Figure\,\ref{fig:param}, where we
display only WDs  with relative error in  their effective temperatures
below 10  per cent (128 objects),  and absolute errors below  0.15 dex
(in \logg  ; 30 objects)  and 0.1  M$_{\odot}$ (in mass;  50 objects).
Moreover, we  only took  into account DA/M  binaries with  available M
dwarf spectral subclasses (318 objects). Visual inspection reveals the
WD parameters are broadly consistent with each other.  Quantitatively,
if we define $\tau$ \citep{Gentile2015MNRAS.452..765G}:
\begin{equation}
\tau = \frac{SDSS\,\mathrm{value} - LAMOST\,\mathrm{value}}{\sqrt{SDSS\,\mathrm{error^2}+LAMOST\,\mathrm{error^2}}}
\end{equation}
then  we find  that  the  WD parameters  show  a comparable  agreement
($\tau<$ 2)  in 70 (\Teff), 77  (\logg) and 82 (mass)  per cent of
the cases. We  can conclude then that the WD  parameters derived using
LAMOST and SDSS spectra are indeed broadly in agreement.

Visual comparison of the M  dwarf spectral subclasses obtained fitting
the    SDSS   and    LAMOST    spectra   (bottom    left   panel    of
Figure\,\ref{fig:param}) seems  to indicate  a large spread  of values
for a  given spectral subtype.  However,  only in 12 of  the 318 cases
considered the difference  in spectral subtypes is larger  than 2, and
in 49 cases larger than 1. We visually inspected the LAMOST spectra of
these 49 objects  and found that they were either  dominated by the WD
contribution and/or of low SN ratio. These are clear cases in which we
expect  higher  uncertainties in  the  determination  of the  M  dwarf
spectral  subclasses. In  order to  quantitatively study  the apparent
disagreement further, we obtained the average LAMOST spectral subclass
and its  standard deviation fixing  the SDSS spectral  subclasses. The
result is shown  as red solid dots  and red solid lines  in the bottom
left panel of Figure\,\ref{fig:param}.  It  becomes clear that, in the
majority of  cases, the spectral  subtypes agree well with  each other
and that the apparent scatter at a given spectral subtype is caused by
isolated cases.

\begin{table}
\centering
\caption{\label{t-ks}
KS and  $\chi^2$ probabilities. Sample  1 refers to the  entire LAMOST
DR5  catalogue excluding  the  SDSS  photometrically selected  objects
observed  by  LAMOST, Sample  2  refers  to the  SDSS  photometrically
selected sample observed by LAMOST, Sample  3 refers to the LAMOST DR1
catalogue of \citet{Ren2014A&A...570A.107R} and Sample 4 refers to the
full LAMOST DR5 catalogue.}
\setlength{\tabcolsep}{1.4ex}
\begin{tabular}{ccc}
\hline
\hline
                    & Sample 1 vs. Sample 2 & Sample 3 vs. Sample 4 \\
\hline
WD \Teff   &     0.09              &         0.35          \\
WD \logg        &     0.66              &         0.46          \\
WD mass             &     0.40              &         0.09          \\
M dwarf Sp          &     0.93              &         0.56          \\
\hline
\end{tabular}
\end{table}

\subsection{Properties of the LAMOST DA/M binary catalogue}

In the  previous section we  determined the stellar parameters  of our
LAMOST  DR5  DA/M  binaries.   This  allows us  to  present  here  the
corresponding  stellar  parameter  distributions,  which  we  show  in
Figure\,\ref{fig:histo} in  the following order: WD  mass (top panel),
WD  effective  temperature  (top-middle  panel),  WD  surface  gravity
(bottom-middle  panel) and  M dwarf  spectral subtype  (bottom panel).
Following  Section\,\ref{s-param},  and  in order  to  exclude  values
subject to large uncertainties, we  exclude WDs with relative error in
their effective temperatures  above 10 per cent (we are  left with 277
objects  in  the  top-middle panel  of  Figure\,\ref{fig:histo}),  and
absolute errors  above 0.15 dex  in \logg\ (76  objects; middle-bottom
panel) and 0.1  M$_{\odot}$ in mass (116 objects; top  panel).  In the
same way, we only consider  DA/M binaries with well-determined M dwarf
spectral     subclasses    (686     objects;    bottom     panel    of
Figure\,\ref{fig:histo}).    Visual   inspection  of   the   parameter
distributions   reveals    that   the   WDs   are    concentrated   at
$\sim$\,0.6\,\Msun\, (which is equivalent to say they are concentrated
at
\logg\ $\sim$\,8\,dex), with a steep decline towards higher and lower
masses  (being  the latter  WDs  that  are  presumably part  of  close
binaries, see Section\,\ref{s-closebin} for  a further discussion) and
with  typical   effective  temperature  values  between   10\,000  and
40\,000\,K (with  a peak at  $\sim$15\,000\,K).  The M  dwarf spectral
types are concentrated between M1--M4,  with a decline towards earlier
and   later    types.    For    completeness,   we   also    show   in
Figure\,\ref{fig:histo} the  parameter distributions that  result from
considering only the photometrically-selected  SDSS DA/M binary sample
that has been observed by LAMOST. We  note that we did not include any
DA/M binary considered as candidate in this exercise.

We run  Kolmogorov-Smirnov (KS) tests  to the WD  cumulative parameter
distributions and  apply a  $\chi^2$ test  to the  cumulative spectral
type  distributions resulting  from the  SDSS photometrically-selected
sample and the total sample of LAMOST DR5 DA/M binaries (excluding the
SDSS photometrically selected ones).  The cumulative distributions are
shown in the  left panels of Figure\,\ref{fig:ks} and  the results are
given in Table\,\ref{t-ks}. The probabilities  obtained do not seem to
indicate the two samples are statistically different, which is what we
should expect given  that the photometric sample is  selected based on
colour cuts that favour the detection of systems containing cooler WDs
and/or earlier type companions. This may  be a consequence of the fact
that it  is more difficult to  derive reliable WD parameters  for this
type of  systems. That is,  cooler WDs are systematically  fainter and
hence  their spectra  are of  lower  SN ratio,  which translates  into
larger  parameter uncertainties.   Hence  the  objects containing  the
coolest WDs  either cannot be fitted  because they are subject  to too
low SN  spectra or the fitted  values do not survive  our quality cuts
and are excluded from the analysis.  This, unfortunately, implies that
our  sample  of  LAMOST  DR5  DA/M  binaries  with  available  stellar
parameters is  still biased  against systems  containing cool
WDs.

We  also run  KS and  $\chi^2$ tests  to the  cumulative distributions
resulting from  the entire LAMOST DR5  DA/M sample and the  LAMOST DR1
catalogue   we  presented   in  \citet{Ren2014A&A...570A.107R}.    The
distributions are  shown in  the right panels  of Figure\,\ref{fig:ks}
and in Table\,\ref{t-ks} we provide  the probabilities. Again, we find
no strong  indications for the  DR1 and  DR5 LAMOST populations  to be
statistically different.

\begin{table*}
\centering
\caption{\label{t-rv}
Radial velocities  measured from  the \Lines{Na}{I}{8183.27,\,8194.81}
absorption  doublet and  the \Ha\  emission  for the  LAMOST DR5  WDMS
binaries. The Heliocentric Julian Dates  (HJD) are also indicated. The
last column lists the telescope used for obtaining the spectra (either
LAMOST or SDSS).  We use `$-$' to indicate that  no radial velocity is
available. The complete  table can be found in  the electronic version
of the paper.}
\setlength{\tabcolsep}{4.4ex}
\begin{small}
\begin{tabular}{ccccccc}
\hline
\hline
Jname & HJD & RV(\Ion{Na}{I}) & Err & RV(\Ha) & Err & Telescope \\ &
(days) & (\kms) & (\kms) & (\kms) & (\kms) & \\
\hline
J000448.23+343627.4 & 2456948.07023 & $-$21.59 & 45.97 & $-$96.14 & 17.32 & LAMOST \\
J000448.23+343627.4 & 2456948.09106 & $-$45.61 & 20.86 & $-$93.60 & 18.83 & LAMOST \\
J000448.23+343627.4 & 2456948.09341 & $-$43.07 & 15.78 & $-$62.27 & 24.84 & LAMOST \\
J000448.23+343627.4 & 2456948.11661 & $-$37.64 & 20.67 & $-$76.21 & 19.09 & LAMOST \\
J000612.32+340358.3 & 2456948.07026 & $-$4.35 & 12.62 & $-$12.92 & 10.50 & LAMOST \\
J000612.32+340358.3 & 2456948.09109 & 39.19 & 12.72 & 14.71 & 10.45 & LAMOST \\
J000612.32+340358.3 & 2456948.09344 & 30.75 & 11.87 & 9.55 & 10.28 & LAMOST \\
J000612.32+340358.3 & 2456948.11664 & 64.64 & 12.63 & 26.05 & 10.43 & LAMOST \\
J000729.32+023124.5 & 2455892.99443 & 14.98 & 14.38 & $-$ & $-$ & LAMOST \\
\hline
\end{tabular}
\end{small}
\end{table*}

\begin{table*}
\centering
\caption{\label{t-newpceb}
The 76 new PCEBs found in this work. The Flag column indicates whether
the PCEBs  are detected based  on \Ion{Na}{I} and/or \Ha\  emission RV
measurements (only from \Ion{Na}{I}: 10; only from \Ha; 01 ; from both
\Ion{Na}{I} and \Ha: 11). The maximum RV shifts are also indicated.}
\setlength{\tabcolsep}{0.5ex}
\begin{small}
\begin{tabular}{cccccccccccc}
\hline
\hline
Jname & Flag & RV shift & RV shift & Jname & Flag & RV shift & RV shift& Jname & Flag & RV shift & RV shift \\
      &      &  (km/s)    & (km/s) &      &       & (km/s)  & (km/s)   &         &    &   (km/s)  & (km/s) \\
      &      & \Ion{Na}{I} & \Ha &         &      & \Ion{Na}{I} & \Ha  &        &     & \Ion{Na}{I} & \Ha  \\
\hline
J000612.32+340358.3   & 10 & 68.99	& $-$	& J093507.99+270049.2   & 11 & 248.88	& 182.41	&   J122514.92+292523.1   & 11 & 113.00		& 87.11	\\ 
J004751.47+340212.7   & 01 & $-$		& 70.77	& J093809.28+143036.9   & 11 & 100.51	& 113.59	&   J123339.40+135943.2   & 11 & 133.43		& 106.79	\\ 
J011739.90+340209.4   & 11 & 177.57	& 158.24	& J094101.91+510719.8   & 01 & $-$		& 59.05	&   J123642.53+580230.6   & 11 & 86.16		& 69.09	\\ 
J011847.16+050531.2   & 11 & 155.83	& 100.75	& J095641.79+013130.7   & 01 & $-$		& 87.01	&   J124831.79+435318.3   & 11 & 78.02		& 121.40	\\ 
J024924.76+071344.3   & 01 & $-$		& 118.56	& J101356.31+272410.8   & 11 & 600.06	& 652.14	&   J124830.76+540803.6   & 11 & 83.12		& 219.67	\\ 
J025301.60$-$013006.8 & 01 & $-$		& 72.39	& J101616.82+310506.5   & 11 & 114.66	& 66.68	&   J132001.24+112805.3   & 10 & 79.95		& $-$	\\ 
J030944.99+285507.5   & 10 & 58.73	& $-$	& J103125.93+020751.4   & 11 & 76.84		& 61.33	&   J132341.89+541636.5   & 10 & 137.35		& $-$	\\
J032750.32+222618.9   & 10 & 109.05	& $-$	& J104318.65+305012.9   & 10 & 146.37	& $-$	&   J134051.00+321015.9   & 01 & $-$			& 127.69	\\
J050825.81+252050.8   & 11 & 102.87	& 73.50	& J104534.46+315740.6   & 01 & $-$		& 80.16	&   J135825.68+171204.1   & 11 & 377.78		& 235.72	\\
J054119.24$-$051838.4 & 11 & 154.71	& 167.96	& J105352.90+340923.5   & 01 & $-$		& 254.71	&   J140115.68+214157.1   & 10 & 49.64		& $-$	\\
J060040.31+312118.6   & 10 & 45.48	& $-$	& J105657.35+330416.2   & 11 & 102.26	& 68.60	&   J142917.86+285103.0   & 10 & 90.38		& $-$	\\
J063840.55+130253.5   & 01 & $-$		& 101.87	& J110827.40+303031.3   & 01 & $-$		& 73.91	&   J144307.83+340523.5   & 01 & $-$			& 94.57	\\
J073128.30+264353.6   & 11 & 56.08	& 161.49	& J111614.48+535720.9   & 01 & $-$		& 74.54	&   J145248.79+234807.6   & 01 & $-$			& 54.60	\\
J081126.70+053912.9   & 01 & $-$		& 293.08	& J112007.64+250221.3   & 11 & 193.70	& 204.91	&   J145430.02+203902.5   & 11 & 120.18		& 148.83	\\
J081654.35+354230.6   & 10 & 146.69	& $-$	& J112124.82+273046.2   & 11 & 256.30	& 252.29	&   J151042.63+273509.9   & 10 & 90.99		& $-$	\\
J082214.50+433343.6   & 10 & 98.77	& $-$	& J113102.81+522645.3   & 01 & $-$		& 74.78	&   J153934.63+092221.5   & 11 & 175.68		& 118.65	\\
J082656.51+480545.7   & 11 & 148.67	& 145.12	& J113910.49+315013.1   & 10 & 121.59	& $-$	&   J153914.69+263710.8   & 01 & $-$			& 290.97	\\
J082845.07+133551.6   & 11 & 227.51	& 210.96	& J114732.48+591735.1   & 10 & 60.98		& $-$	&   J154101.07+294829.4   & 11 & 168.15		& 185.74	\\
J083056.11+315941.9   & 01 & $-$		& 69.67	& J114732.25+593921.9   & 10 & 106.47	& $-$	&   J160645.02+284725.9   & 11 & 88.25		& 100.14	\\
J084418.12+340458.9   & 11 & 122.49	& 105.03	& J115057.60+282637.8   & 11 & 235.92	& 160.12	&   J165354.16+322937.6   & 01 & $-$			& 75.64	\\
J085024.05+054757.8   & 10 & 37.21	& $-$	& J115026.28+302359.4   & 01 & $-$		& 231.17	&   J213136.72+002947.6   & 10 & 74.41		& $-$	\\
J085025.75+344053.2   & 10 & 98.86	& $-$	& J115113.14$-$011111.1 & 01 & $-$		& 168.85	&   J224522.49+063817.2   & 01 & $-$			& 88.17	\\
J085414.26+211148.1   & 11 & 233.93	& 149.53	& J115510.86+271324.3   & 01 & $-$		& 95.53	&   J224805.92+144328.3   & 01 & $-$			& 66.04	\\
J085634.83+373913.4   & 11 & 180.94	& 202.61	& J121301.82+282310.0   & 11 & 389.86	& 388.75	&   J234106.82+083550.3   & 01 & $-$			& 81.32	\\
J090210.95+252913.3   & 01 & $-$		& 114.09	& J122037.01+492334.0   & 10 & 76.86		& $-$	&   J235408.04+291623.1   & 01 & $-$			& 54.65	\\
J092712.02+284629.2   & 01 & $-$		& 96.31	&                       &    & & &                         &    & & \\
\hline
\end{tabular}
\end{small}
\end{table*}

\begin{figure}
  \begin{center}                             \includegraphics[angle=0,
  width=\columnwidth]{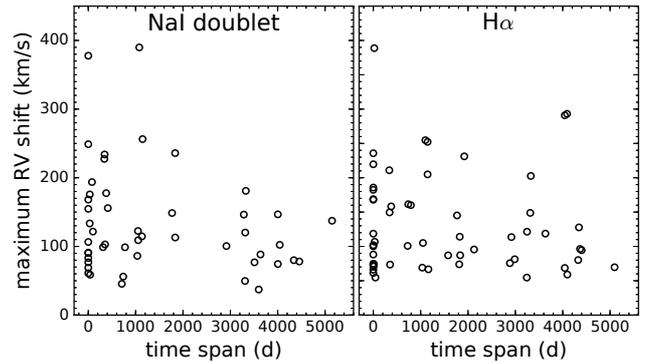} \caption{\label{fig:rv-time} Left
  panel: The maximum  \Ion{Na}{I} doublet RV shift  measured vs.  time
  span between  the observed spectra  (in days)  for the 76  new PCEBs
  identified   in   this   work.     Right   panel:   The   same   but
  for  \Ha\,emission.   The numerical  values  of  the RV  shifts  are
  provided in Table\,\ref{t-newpceb}.}  \end{center}
\end{figure}

\section{Identification of close WDMS binaries}
\label{s-closebin}

In  this section  we aim  at  detecting PCEBs  among the  WDMS
binaries identified  in this work.   To that end we  require measuring
the radial velocities (RVs) of at least one stellar component.  WD RVs
can        be        measured         fitting        the        Balmer
lines      \citep{Breedt2017MNRAS.468.2910B}     and/or      employing
cross-correlation    techniques   \citep{Anguiano2017MNRAS.469.2102A}.
However, we expect the WD RV  shifts to be smaller than those measured
from their M  dwarf companions, since the WDs are  generally closer to
the  center of  mass  in this  type of  binaries.   We hence  measure
the \Lines{Na}{I}{8183.27,\,8194.81}  absorption doublet and  the \Ha\
emission RVs arising from the M dwarf companions from all the
LAMOST WDMS  binary spectra  identified in this  work.  We  identify a
PCEB when we detect more  than 3$\sigma$ RV variation.  Conversely, if
no RV variation  is detected from spectra taken separated  by at least
one night, the system is considered  as a likely wide binary candidate
that presumably  evolved avoiding episodes  of mass transfer.   We fit
the   \Lines{Na}{I}{8183.27,\,8194.81}  absorption   doublet  with   a
second-order polynomial  plus a double-Gaussian line  profile of fixed
separation, while the \Ha\ emission  line (if present in the spectrum)
is fitted with  a second-order polynomial plus  a single-Gaussian line
profile                             \citep{Rebassa2008MNRAS.390.1635R,
Ren2013AJ....146...82R}. With the aim of  increasing the SN ratio, the
final released LAMOST DR5 spectra  are the result of combining several
different  sub-exposure spectra  (hereafter  sub-spectra).  Hence,  we
measure the  RVs from all  available sub-spectra  as well as  from the
final combined spectra.

\begin{figure}
  \begin{center} \includegraphics[angle=-90,
    width=\columnwidth]{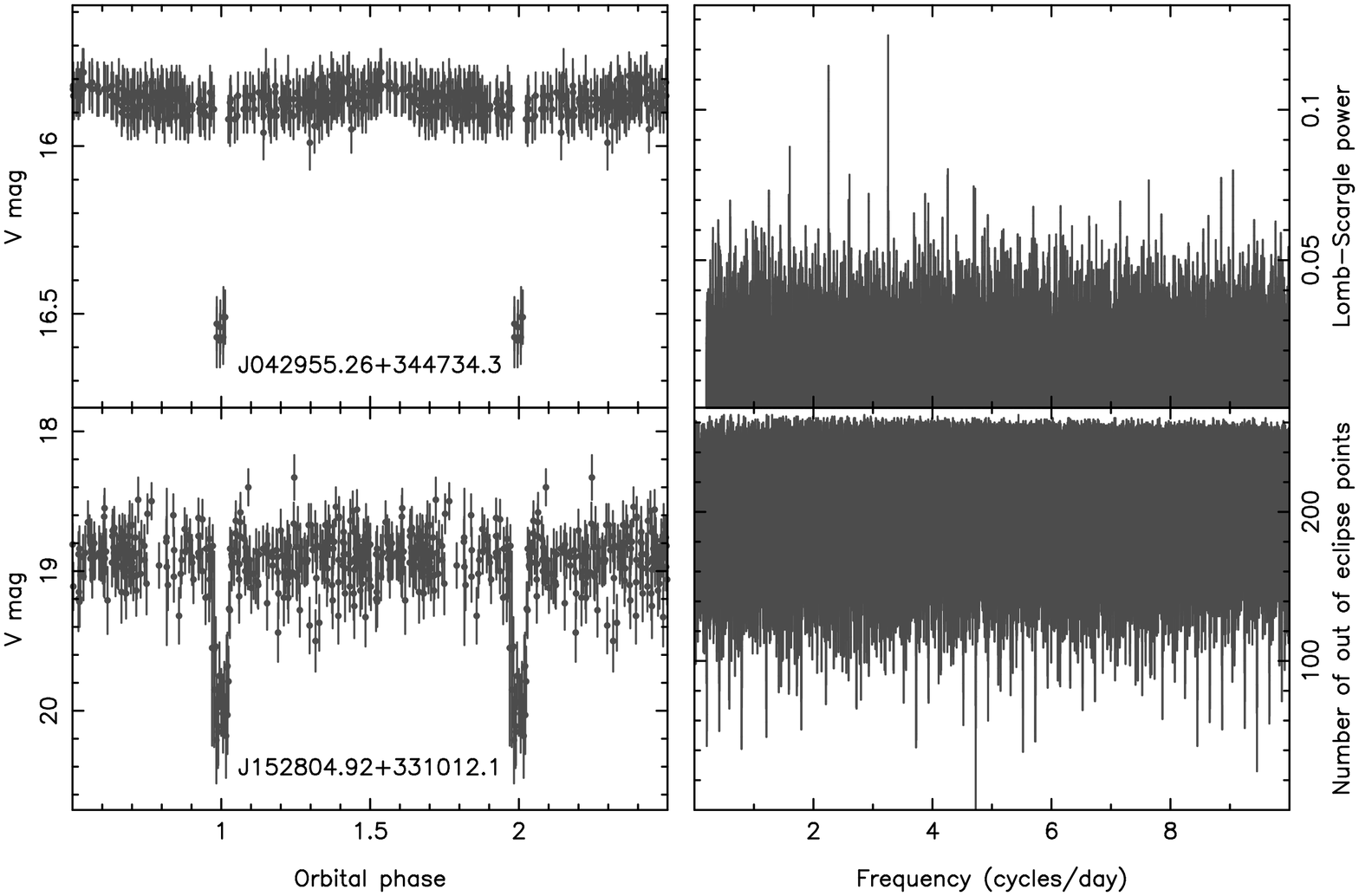} \caption{\label{fig:eclip} Left
    panels: CSS light curves folded over the best orbital period
    determinations (see right panels) of the two new eclipsing DA/M
    binaries identified in this work.  Top right panel: the results of
    a standard Lomb-Scargle periodogram applied to the CSS photometry
    of J042955.26+344734.3, peaking at 3.25556 cycles/day
    (corresponding to a period of 0.30716\,d). Bottom right panel:
    displayed are the results obtained applying a box-search method
    (see Section\,\ref{s-eclip} for details) to the CSS photometry of
    J152804.92+331012.1, resulting in an orbital period of 0.21155
    days.}  \end{center}
\end{figure}

\begin{table*}
\centering
\caption{\label{t-eclip}
LAMOST  DR5  WDMS binaries  that  display  light curve  variations  as
revealed  by the  Catalina photometric  surveys. We  provide notes  on
individual  systems, classification  and orbital  periods obtained  by
calculating periodograms from the photometric data.}
\setlength{\tabcolsep}{4.4ex}
\begin{small}
\begin{tabular}{cccc}
\hline
\hline
Object &  Notes & Classification  & Orbital Period \\
\hline
J000729.32+023124.5 & Some faint points  & Candidate new eclipser &      ? \\
J001752.63+332424.9 & Reflection effect? & Candidate non-eclipser &   6.4 hours? \\
J003033.11+070657.5 & Seems to display variation & Candidate non-eclipser & 1.2 hours? or double \\
J004232.56+415403.0 & Reflection effect? & Non-eclipser & 0.10312 days \\
J004751.47+340212.7 & Large variations & Candidate non-eclipser &  ? \\
J005130.56+314555.9 & Reflection effect? & Candidate non-eclipser & 1.7 hours? \\
J010010.53+003739.1 & Reflection effect? & Candidate non-eclipser & 1.7 hours? \\
J011547.58+005350.0 & Not many points & Candidate non-eclipser & 1.2 hours? \\
J012549.89+330940.6 & Variations & Non-eclipser & 0.24244 days, or double \\
J012550.48+280756.0 & Some faint points & Candidate new eclipser & ? \\
J013157.96+084948.2 & Reflection effect? & Candidate non-eclipser & 2.3 hours? \\
J024924.77+071344.3 & Variations & Non-eclipser & 0.17334 days, or double \\
J030308.35+005444.1 & Clear eclipser & Known eclipser [1] & 0.13444 days \\
J042955.26+344734.3 & Clear eclipser & New eclipser & 0.30716 days \\
J064959.81+424110.2 & Reflection effect? & Non-eclipser & 0.24496 days \\
J082145.27+455923.4 & Clear eclipser & Known eclipser [2] & 0.50909 days \\
J084028.85+501238.2 & Variation & Candidate non-eclipser & 1.2 hours? or double \\
J085414.28+211148.2 & Reflection effect? & Non-eclipser & 0.10215 days \\
J085835.56+281356.3 & large variations & LARP? & ? \\
J090812.04+060421.2 & Clear eclipser & Known eclipser [3] & 0.14944 days \\
J091216.37+234442.5 & Reflection effect? & Non-eclipser & 0.26356 days \\
J092712.02+284629.2 & Variation & Candidate non-eclipser & 3.1 hours? or double \\
J092741.73+332959.1 & Clear eclipser & Known eclipser [2] & 2.30822 days \\
J093207.63+334805.9 & Some faint points & Candidate new eclipser & 1.7 days? \\
J093507.99+270049.2 & Clear eclipser & Known eclipser [4] & 0.20103 days \\
J093947.95+325807.3 & Clear eclipser & Known eclipser [3] & 0.33098 days \\
J094913.36+032254.5 & Reflection effect? & Candidate non-eclipser & 3.1 hours? \\
J095719.24+234240.7 & Clear eclipser & Known eclipser [3] & 0.15087 days \\
J095737.59+300136.5 & Clear eclipser & Known eclipser [2] & 1.92613 days \\
J101307.79+245713.1 & Clear eclipser & Known eclipser [5] & 0.12904 days \\
J104012.99+252559.9 & Reflection effect? & Candidate non-eclipser & 3.7 hours? \\
J110827.40+303031.3 & Variation & Non-eclipser & 0.81820 days, or double \\
J112007.64+250221.3 & Some faint points & Candidate new eclipser &  ? \\
J112738.71+281532.7 & Several faint points & Candidate new eclipser & ? \\
J113102.81+522645.3 & Variation & Candidate non-eclipser & 2 hours? or double \\
J114224.71$-$022610.0 & Variation & Non-eclipser & 0.56112 days, or double \\
J114509.77+381329.2 & Variation & Non-eclipser & 0.19004 days, or double \\
J114853.34+555217.0 & Active star? Flares? & Active M dwarf? & ? \\
J120020.81+363557.3 & Some faint points & Candidate new eclipser & ? \\
J121010.13+334722.9 & Clear eclipser & Known eclipser [6] & 0.12449 days \\
J121258.25$-$012309.9 & Clear eclipser & Known eclipser [7] & 0.33587 days \\
J122630.86+303852.5 & Reflection effect? & Non-eclipser & 0.25869 days \\
J123214.38+351324.8 & Variation & Candidate non-eclipser & 2.5 hours? \\
J132925.21+123025.4 & Clear eclipser & Known eclipser [2] & 0.08097 days \\
J134234.78+304849.2 & Several faint points & Candidate new eclipser & 1.66 days? \\
J135825.68+171204.1 & Variation & Non-eclipser & 0.16046 days, or double \\
J141811.97+204150.8 & Reflection effect? & Candidate non-eclipser & 2.6 hours? \\
J143547.87+373338.5 & Clear eclipser & Known eclipser [8] & 0.12563 days \\
J143900.62+560219.0 & Variation & Non-eclipser & 0.34268 days, or double \\
J144307.83+340523.5 & Shows big variations & Unclear & ? \\
J144846.85+071304.3 & Some faint points, real? & Candidate new eclipser & ? \\
J151426.90+285720.4 & Flares? CV? & Unclear & ? \\
J152804.92+331012.1 & Clear eclipser & New eclipser & 0.21155 days \\
J154846.00+405728.7 & Clear eclipser & Known eclipser [8] & 0.18552 days \\
J212309.40+040929.5 & Many faint points & Candidate new eclipser & ? \\
J212531.92$-$010745.9 & Reflection effect & Non-eclipser & 0.28982 days \\
J233900.38+115707.2 & Variation & Non-eclipser & 0.12286 days, or double \\
\hline
\end{tabular}
\end{small}
\begin{minipage}{\textwidth}
[1] \citet{Parsons2013MNRAS.436..241P};
[2] \citet{Parsons2013MNRAS.429..256P}; [3] \citet{Drake10arxiv};
[4] \citet{Drake14}; [5] \citet{Parsons2015MNRAS.449.2194P};
[6] \citet{Pyrzas2012MNRAS.419..817P};
[7] \citet{Parsons2012MNRAS.420.3281P};
[8] \citet{Pyrzas2009MNRAS.394..978P}
\end{minipage}
\end{table*}

We  cross-matched our  LAMOST DR5  catalogue with  the current  newest
spectroscopic data  base of  SDSS, i.e.  DR14.   This resulted  in 685
additional spectra  for 465 of  our targets. Combining LAMOST  DR5 and
SDSS DR14 we thus gathered a  total of 1835 spectra (1150 from LAMOST,
685   from   SDSS)  for   measuring   RVs.    We  derived   at   least
one \Lines{Na}{I}{8183.27,\,8194.81}  absorption doublet and  one \Ha\
emission RV of an  accuracy better than 20 \kms\ for  685 and 477 WDMS
binaries in our catalogue, respectively. For 607 and 408 WDMS binaries
we  managed  to  measure  at  least  two  \Ion{Na}{I}  and  \Ha\  RVs,
respectively.   The   RVs  are  provided  in   Table\,\ref{t-rv}  (for
completeness,  we also  provide the  RVs we  measured with  accuracies
worse than 20\,\kms).

We   identified   89   and    103   objects   displaying   more   than
3$\sigma$ \Ion{Na}{I} and \Ha\ RV variation, respectively.  If we only
take  into account  systems with  RVs taken  on different  nights (337
objects  for \Ion{Na}{I},  229  objects  for \Ha),  we  obtain a  PCEB
fraction  of   $\sim$26  per   cent  based   on  the   \Ion{Na}{I}  RV
measurements,  or   $\sim$  45   per\,cent  based   on  the   \Ha\  RV
measurements.   The  PCEB fraction  derived  from  the \Ion{Na}{I}  RV
analysis agrees  well with those  obtained in previous works  from the
SDSS     WDMS      sample     \citep[e.g.][]{Nebot2011A&A...536A..43N,
Rebassa2016MNRAS.458.3808R}. It has  to be noted that  the higher PCEB
fraction derived  by analysing  the \Ha\  RVs needs  to be  taken with
caution.  This  is because \Ha\ RVs  are less robust for  detecting RV
variations          due          to         magnetic          activity
effects \citep{Rebassa2008MNRAS.390.1635R}.

The total number of unique PCEBs  we have identified that results from
excluding duplicated targets  from both the \Ion{Na}{I}  and \Ha\ PCEB
lists  is   128.   Among   these,  four  were   previously  identified
by \citet{Ren2014A&A...570A.107R} within the  LAMOST DR1 catalogue and
50 by  \citet{Rebassa2016MNRAS.458.3808R} within the SDSS  DR12 sample
(we note  that two were both  detected within the LAMOST  DR1 and SDSS
DR12 catalogues).  Thus, we are  left with 76 new PCEB identifications
in this work.   In table\,\ref{t-newpceb} we provide  the object names
of  these 76  PCEBs and  in Figure  \ref{fig:rv-time} we  show
their  maximum  radial  velocity  shift vs.   time  span  between  the
observed  spectra. It  is worth  mentioning that  in $\sim$1/2  of the
cases the time span  is over 1000 days.  This is due  to the fact that
we  combine LAMOST  and SDSS  spectra to  detect close  binaries, thus
resulting in time baselines as long as $\sim$15 years.

\section{Eclipsing WDMS systems}
\label{s-eclip}

Following the approach presented in \citet{Parsons2013MNRAS.429..256P}
and  \citet{Parsons2015MNRAS.449.2194P}, we  cross-matched our  LAMOST
DR5 WDMS binary catalogue with  the photometric data from the Catalina
Sky    Survey    (CSS)    and    Catalina    Real    Time    Transient
Survey    \citep[CRTS][]{Drake2009ApJ...696..870D}.    This    allowed
detecting WDMS binaries displaying light curve variations. During this
process the  raw CSS  data were  re-reduced by  us to  identify deeply
eclipsing systems and to remove contaminated exposures.

Among the 876 unique WDMS binaries that form the DR5 LAMOST catalogue,
687  were observed  by  the Catalina  surveys, of  which  630 did  not
display light  curve variations. The  remaining 57 objects  include 16
eclipsing  systems, two  of which  are new  (see the  light curves  in
Figure\,\ref{fig:eclip}),  9  new  eclipsing  candidates,  14  objects
displaying reflection  effects or  irradiation effects,  14 candidates
for  displaying  reflection  or  irradiation  effects,  one  candidate
low-accretion rate polar, one candidate active M dwarf and two systems
displaying light curve  variations which we are not  able to classify.
We  provide  additional notes  and  our  classification for  these  57
objects in Table\,\ref{t-eclip}.

We  calculated   \citet{Scargle82}  periodograms  from   the  Catalina
photometric data with the aim of  measuring the orbital periods of the
57  objects  displaying light  curve  variations.   This is  a
suitable method when analysing  light curves displaying out-of-eclipse
variations due to reflection or ellipsoidal modulation effects (see an
example in  the top right  panel of Figure\,\ref{fig:eclip}).   In the
absence of such variations, i.e.  only the eclipses are sampled by the
CSS   data,   we   employed    the   box   fitting   method   outlined
by   \citet{Parsons2013MNRAS.429..256P}.    We  first   identify   the
in-eclipse points  from the raw light  curve and fold the  data over a
given orbital period.  From the folded  data we identify the first and
last in-eclipse points  and count the number  of out-of-eclipse points
in between. This  procedure is repeated over a large  range of adopted
orbital  periods,  being  the  correct determination  the  value  that
results in  zero out-of-eclipse  points. The number  of out-of-eclipse
points  as a  function of  the adopted  orbital period  for our  newly
identified eclipsing system J152804.92+331012.1 is shown in the bottom
right panel of Figure\,\ref{fig:eclip}.

For 30 systems we were able to determine precise values of the orbital
periods. In additional 15 cases, we could only derive estimates of the
orbital  periods  due  to  insufficient information  provided  by  the
periodograms. We note that since, a priory, we do not know whether the
light curve  variations are due  to reflection of  irradiation effects
for our non-eclipsing systems, there is a possibility that the orbital
periods are double the measured  ones.  We include our measured values
of the orbital periods in Table\,\ref{t-eclip}.  We note that, for the
two  new eclipsing  systems found  here, we  were not  able to  derive
accurate RVs due to  the low SN ratio of their  spectra, thus they are
not included in our PCEB list.

\section{Summary and conclusions}

The catalogue  of WDMS binaries  from LAMOST DR5 contains  876 objects
and it  is $\sim$8 times larger  than our previous DR1  sample. 357 of
these  systems   ($\sim$40  per  cent   of  the  catalogue)   are  new
identifications  that have  not been  published before.  Moreover, 339
were  observed as  part of  a  dedicated LAMOST  survey for  obtaining
spectra of WDMS binaries photometrically selected within SDSS that are
expected to contain cool WDs and/or early type M dwarf companions.

We   determined  the   stellar  parameters   (white  dwarf   effective
temperatures,  surface  gravities and  masses,  and  M dwarf  spectral
types) of  our systems following a  decomposition/fitting routine, and
we  used  the corresponding  parameter  distributions  to analyse  the
intrinsic  properties of  the LAMOST  DR5  sample. We  found that  the
population of cool WDs remains under-represented.  This is most likely
due to the fact that  cool (\Teff$\la$10\,000 K) WDs are systematically
fainter and hence associated to lower SN ratio spectra.  This increases
considerably the  probability for our decomposition/fitting  method to
determine WD  parameters associated to  large uncertainties and,  as a
consequence, these objects are not taken into account in our analysis.

We  measured the  \Lines{Na}{I}{8183.27,\,8194.81} absorption  doublet
and  the  \Ha\  emission  radial  velocities of  each  object  in  our
catalogue from  their LAMOST DR5  as well  as SDSS DR14  spectra (when
available).   We   detected  128   systems  (76   of  which   are  new
identifications)  displaying  more   than  3$\sigma$  radial  velocity
variations  and hence  classify  these objects  as  PCEBs.  The  close
binary fraction  we derived  is $\sim$26 per  cent, in  agreement with
previous studies.

By cross-matching our catalogue with  the Catalina Surveys we found 57
systems displayed light curve variations. Among these we identified 16
eclipsing systems, two of which are new, and nine additional eclipsing
WDMS binary candidates.  By analysing the periodograms calculated from
the photometric data we were able  to determine the orbital periods of
30 objects and estimate the orbital periods of 15 additional systems.

\section*{Acknowledgements}

We  thank  the   anonymous  referee  for  the   helpful  comments  and
suggestions. This  work is  supported by Joint  Funds of  the National
Natural  Science Foundation  of  China (Grant  No.  U1531244) and  the
National  Key  Basic  Research  Program  of  China  2014CB845700.  JJR
acknowledges  support  from the  Young  Researcher  Grant of  National
Astronomical   Observatories,  Chinese   Academy   of  Sciences.   ARM
acknowledges  support  from  the  MINECO under  the  Ram\'on  y  Cajal
programme  (RYC-2016-20254) and  the  AYA2017-86274-P  grant, and  the
AGAUR grant SGR-661/2017.

This work has made use of data products from the Guoshoujing Telescope
(the  Large  Sky  Area  Multi-Object  Fibre  Spectroscopic  Telescope,
LAMOST). LAMOST  is a National  Major Scientific Project built  by the
Chinese Academy of Sciences. Funding for the project has been provided
by the National Development and  Reform Commission. LAMOST is operated
and  managed  by  the  National  Astronomical  Observatories,  Chinese
Academy of Sciences.

%%%%%%%%%%%%%%%%%%%%%%%%%%%%%%%%%%%%%%%%%%%%%%%%%%

%%%%%%%%%%%%%%%%%%%% REFERENCES %%%%%%%%%%%%%%%%%%

%\bibliographystyle{mnras}
%\bibliography{dr5wdms}

%%%%%%%%%%%%%%%%%%%%%%%%%%%%%%%%%%%%%%%%%%%%%%%%%%

%%%%%%%%%%%%%%%%% APPENDICES %%%%%%%%%%%%%%%%%%%%%

%\appendix

%\section{Appendix}

%%%%%%%%%%%%%%%%%%%%%%%%%%%%%%%%%%%%%%%%%%%%%%%%%%

% Don't change these lines
\bsp	% typesetting comment
\label{lastpage}
\end{document}